\documentclass[
twocolumn,
superscriptaddress,
 amsmath,amssymb,
 aps,
prb,
]{revtex4-2}

\usepackage{graphicx}
\usepackage{dcolumn}
\usepackage{bm}
\usepackage{color}
\usepackage{multirow}
\usepackage{amsmath}

\begin{document}

\title{Phonon anomalies and  critical scaling in the spin-$1/2$ trimer chain Na$_2$Cu$_3$Ge$_4$O$_{12}$}

\author {P. Srikanth Patnaik}
\affiliation {Department  of  Physics, Indian Institute of Technology Kharagpur, Kharagpur 721302, India}

\author {A. K. Bera}
\affiliation {Solid State Physics Division, Bhabha Atomic Research Centre, Mumbai 400085, India}
\affiliation {Homi Bhabha National Institute, Anushaktinagar, Mumbai 400094, India}

\author {Srishti Bhardwaj}
\affiliation {Department of Physics, Indian Institute of Technology Roorkee, Uttarakhand 247667, India}

\author {Tulika~Maitra}
\affiliation {Department of Physics, Indian Institute of Technology Roorkee, Uttarakhand 247667, India}

\author {Buddhananda Banerjee}
\affiliation {Department of Mathematics, Indian Institute of Technology Kharagpur, Kharagpur 721302, India}

\author {Anushree Roy}
\email {anushree@phy.iitkgp.ac.in}
\affiliation {Department  of  Physics, Indian Institute of Technology Kharagpur, Kharagpur 721302, India}

\author {S.M. Yusuf}
\email {smyusuf@barc.gov.in}
\affiliation {Solid State Physics Division, Bhabha Atomic Research Centre, Mumbai 400085, India}
\affiliation {Homi Bhabha National Institute, Anushaktinagar, Mumbai 400094, India}
\affiliation{ \footnotesize UM-DAE Centre for Excellence in Basic Sciences, Vidyanagari, University of Mumbai, Mumbai 400098, India.}

\begin{abstract}
Low-dimensional quantum magnets provide an ideal platform to explore spin-lattice coupling-mediated quantum correlations, which give rise to emergent quasiparticle excitations.
 The antiferromagnetically coupled spin-1/2 trimer chain of copper ions in Na$_2$Cu$_3$Ge$_4$O$_{12}$
(NCGO) hosts high-energy spin excitations of different species, whose energy scales overlap with those of lattice vibrations. Here, we report a comprehensive temperature-dependent Raman spectroscopic study performed between 80 and 400 K. The dynamic spin susceptibility, as obtained from the analysis of the broad spectral background, reveals the emergence of quasiparticle excitations below 170 K.
We further identify  an unusual crossover of phonon dynamics when the material transits from a normal paramagnetic state to a correlated quantum magnetic state. A power law dependence of the integrated  Raman
susceptibility of the phonon modes,  $I_{\chi^{\prime\prime}}^{i}\sim|T-T_{c}|^\beta$, is observed with the critical temperature $T_c$=167$\pm$1 K, and critical exponent $\beta = 0.24\pm 0.02$.  The combined results obtained from the broad spectral background and sharp phonon peaks further indicate that the phonon renormalization observed across the crossover is driven by dynamic spin states. Additionally, statistical correlations among phonon energy eigenvalues, quantified through matrix-norm and power-test analyses of 200 spectra recorded at 80 K, reveal an unexpected linear correlation among phonon modes, also indicating that the collective lattice response is governed by spin correlations. These findings establish NCGO as a model system for investigating cooperative spin–lattice coupling and critical scaling behavior of phonon dynamics in low-dimensional magnetic materials.
\end{abstract}

\maketitle

\section{Introduction}\label{sec1}
The pursuit of high-energy spin excitations in real systems stands as a critical frontier in contemporary condensed matter physics, promising to unravel new insights and deepen our understanding of the physics of complex correlated materials. In such systems, the strong interplay among localized spins, charge, and lattice degrees of freedom gives rise to unconventional ground states and excitations. Understanding the microscopic mechanisms that govern these correlated quantum behaviors remains challenging, as they often defy our conventional theoretical knowledge in the field.   

In this context, low-dimensional systems such as spin chains, ladders, and trimer networks provide an ideal platform for exploring emergent excitations and spin-lattice entanglement under well-defined geometrical constraints. Among these promising candidates, antiferromagnetically coupled spin-$1/2$ trimer chains have emerged as a fascinating system \cite{Cheng2022,Bera2022,Li2025}, hosting three different species of fermionic spin excitations: namely spinons, doublons, and quartons \cite{Cheng2022,Bera2022,cheng2024,prabhakar2025}, whose collective dynamics reflect the underlying quantum coherence of the system. The emergent underlying physics of such systems has attracted attention in recent years, owing to their technological potential for quantum computation \cite{gaita2019,moreno2021}, spin-based information storage \cite{wolf2001,han2018,puebla2020,guo2024}, and quantum sensing \cite{degen2017,aslam2023,bernier2025}. 

\begin{figure}[t]
\centering
\includegraphics[width=\linewidth]{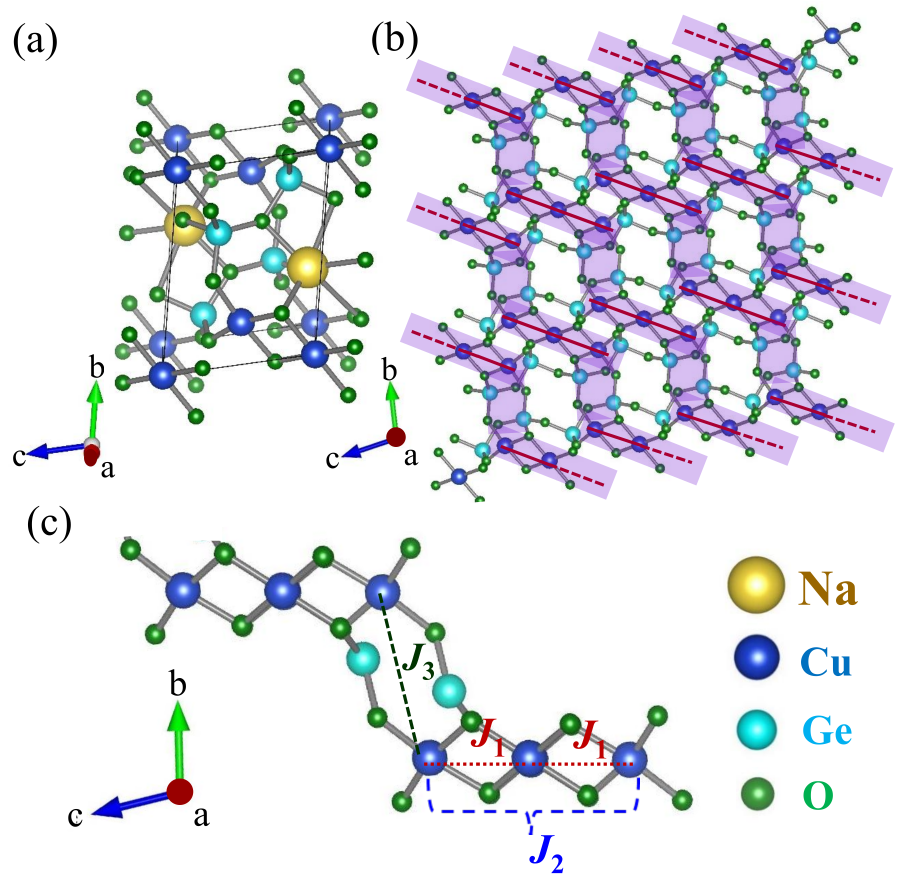} 
\caption{(a)  Unit cell of NCGO with the space group P-1.
The sodium, copper, germanium, and oxygen atoms are shown in yellow, blue, cyan, and green color balls, respectively.
(b) Trimer chain in NCGO with 12 unit cells. Na ions are not shown to avoid cluttering. 
The linear arrangement of three Cu ions forming a trimer of spin-1/2 (shown by red lines) is coupled, forming a zigzag chain (shown as shades) through two Ge$_1$O$_4$ tetrahedra.
Interchain interaction is weak and can be neglected \cite{Bera2022}. 
(c) Chain of two trimers. Nearest neighbor intra-trimer, next-nearest neighbor intra-trimer, and inter-trimer exchange couplings through $J_1$, $J_2$, and $J_3$, respectively, are marked in a linear chain of Cu ions in NCGO. For representations, we used the software VESTA (version 3, 2006–2014) \cite{vesta}.} 
\label {structure}
\end{figure}

The experimental realization of a weakly coupled antiferromagnetic trimer chain is demonstrated in Na$_2$Cu$_3$Ge$_4$O$_{12}$
(NCGO) \cite{Bera2022}. The arrangement of Cu ions in the crystal structure of NCGO, shown in Fig. \ref{structure}(a), leads to a unique spin arrangement. The spin-$1/2$ trimer chain of Cu ions in NCGO is realized by a linear arrangement of edge-sharing three CuO$_4$ square planes, which form periodic arrays of Cu$_3$O$_8$ trimers, as shown by the red lines in Fig. \ref{structure}(b). These trimers are coupled, forming a zigzag chain
(shown by purple shade in Fig. \ref{structure}(b)) through two Ge$_1$O$_4$ tetrahedra \cite{Bera2022}. Each Cu ion of the trimers has one unpaired spin. 
In an isolated trimer, three Cu ions are connected by the nearest-neighbor intra-trimer exchange interactions of the coupling constant $J_1$. They are also coupled with the second nearest neighbor Cu ions, with an interaction coupling constant of $J_2$. The trimers in the chain are coupled by inter-trimer exchange interactions with the coupling constant $J_3$.  Antiferromagnetic exchange coupling schemes through $J_1$, 
$J_2$, and $J_3$ in the linear chain of Cu ions in NCGO are shown in Fig. \ref{structure}(c).
The evidence of emergent spin excitations below 250 K in this compound, NCGO, has been detected from inelastic neutron scattering (INS) measurements by probing the dynamical structure factor, 
corroborated by the calculated spin structure factor using the density matrix
renormalization group approach \cite{Bera2022}. At 3 K, the three non-similar species of spin excitations, spinons, doublons, and quartons, are in the energy scales $\leq$ 5 meV, over 17 meV to 22 meV, and from 32 meV to 37 meV, respectively. The study \cite{Bera2022} further revealed that the low-energy spin excitations in NCGO are dispersive in nature, whereas the intermediate and high-energy spin excitations are gapped and weakly dispersive. 

The unusual spin dynamics in NCGO have been further reported \cite{Bera2022}. 
The temperature-dependent susceptibility, $\chi (T)$ vs $T$ plot,  shows a deviation from normal paramagnetic behavior below 300 K and exhibits a broad maximum around 11 K. The later has been attributed to short-range antiferromagnetic ordering. Upon further lowering the temperature to 2 K, long-range antiferromagnetic interactions set in. In addition, the pulse field magnetization curves show the presence of a 1/3 magnetization plateau; argued as a characteristic feature of weakly antiferromagnetically coupled highly entangled spin states. 

The coupling between lattice vibrations and spin excitations in quantum materials stands out as one of the key aspects that govern the cooperative dynamics of correlated quantum systems. A large number of reports in the literature demonstrate evidence of the coupling of phonons with magnon or multimagnon excitations \cite{streib2019,cui2023,kang2023phonon,sun2025} and fractional excitations in 1D \cite{gnezdilov2012} and 2D \cite{lemmens2000} antiferromagnetic chains, in particular.
Experimental manifestations of the crosstalk between these spin excitations–lattice dynamics have been revealed through various spectroscopic techniques, including neutron scattering \cite{msika2023}, Raman spectroscopy \cite{misochko1996,mai2019,sethi2019,Pal2021}, and resonant inelastic X-ray scattering (RIXS) \cite{Monney2013}. The polarization-resolved Raman studies on FePSe$_3$ identified the signatures of chiral phonon-magnon coupling \cite{cui2023}, while the temperature-dependent studies establish the interplay between magnon and lattice vibrations in the system \cite{sun2025}. These studies reveal the transfer of angular momentum between phonon and magnon \cite{sun2025}.

Temperature-dependent Raman measurements on 1D spin-ladder cuprate systems such as Sr$_2$CuO$_3$ and SrCuO$_2$ provide evidence for free spinon excitations \cite{misochko1996}. It is shown that phonons in antiferromagnetic spin-chain cuprates act as scattering centers \cite{msika2023,hlubek2012} or defects/impurities \cite{chernyshev2016} for the spin excitations, reducing the mean free path of spinons.  
Various theoretical models have been proposed to understand the microscopic mechanisms underlying spinon-phonon interactions in quantum spin liquids from symmetry considerations \cite{serbyn2013,zhang2021}.

From the above studies, it is, therefore, evident that Raman spectroscopy offers a sensitive probe for detecting spin–lattice coupling in low-dimensional quantum magnets, where competing spin-spin exchange pathways and atomic vibrations conspire to generate emergent quasiparticle excitations. In this context, the near-degeneracy of the energy scales of emergent spin excitations and lattice dynamics in NCGO offers a unique platform for investigating the dynamic coupling between spin and lattice degrees of freedom, allowing Raman scattering to simultaneously access both phononic and magnetic excitation channels. Consequently, the temperature-dependent evolution of the Raman spectra encodes valuable information about the crossover from the normal paramagnetic state to the emergent quantum magnetic phase, providing a window into the cooperative nature of spin–lattice dynamics in this trimer-chain quantum system.

\section{Methods}

Polycrystalline NCGO samples were synthesized by the solid-state reaction method by mixing Na$_2$CO$_3$, CuO and GeO$_2$ in a 1:3:4 ratio \cite{Bera2022}. The mixture was annealed for 40 hrs at 1073 K in a furnace, while grinding it several times in between. The sample preparation routes of K$_2$Cu$_3$Ge$_4$O$_{12}$ (KCGO), Li$_2$Ni$_3$P$_4$O$_{14}$ (LNPO), and pseudo-bilayer La$_{1.4}$Sr$_{1.6}$Mn$_2$O$_7$ (BL-LSMO) are reported elsewhere \cite{chikara2023,kumar2021,Chikara2025}.

Micro-Raman measurements of the samples were carried out using the Raman spectrometer of model LabRam HR Evolution (Horiba, France). The spectrometer is equipped with a confocal microscope (BX41, Olympus, Japan), a Peltier-cooled charge-coupled device (CCD) detector (Model Syncerity, USA), and a 532 nm Nd-Yag diode laser as the excitation source. All measurements were performed using an objective lens  50$\times$L to focus the laser beam on the sample and to collect the scattered light in the backscattered geometry. The measurements were carried out for different incident laser powers and integration times to test and avoid unintentional laser heating of the sample during the measurements (See Supplementary Material SM1 \cite{suppli}). All reported data were recorded with 1.25 mW laser power on the sample. Temperature-dependent Raman measurements between 80 K and 400 K were recorded by keeping the sample in a temperature stage (THMS-600, Linkam, UK) equipped with a liquid nitrogen pump and a temperature controller. Temperature-dependent Raman spectra were also recorded using the same spectrometer and the same temperature stage with 568 nm (Ar$^+$-Kr$^+$ gas laser) as the excitation wavelength.

For statistical analysis, we have recorded 200 Raman spectra of all samples using  a triple monochromator spectrometer (T64000, Horiba, France) in subtractive mode. The spectrometer is equipped with an optical
microscope (BX41, Olympus, Japan) along with an objective lens 50×L (numerical aperture 0.50), a Peltier cooled CCD detector (Synapse, Horiba, France). For a 532 nm excitation wavelength, with an 1800 grooves/mm grating and a 1 inch CCD detector,  the resolution of the instrument is approximately 0.7 cm$^{-1}$ ($\sim$ 0.08 meV).

 We performed first-principles calculations using density functional theory (DFT) as implemented in the Vienna Ab Initio Simulation Package (VASP) \cite{kresse1993}. The projector augmented wave (PAW) method was used to deal with the core electrons. As the exchange-correlation functional, we used the Perdew-Burke- Ernzerhof (PBE) parametrization \cite{perdew1996} of the generalized-gradient approximation (GGA). The energy cut-off for the plane wave basis was set at 650 eV.  The Hellmann-Feynman forces convergence criterion was 0.001 eV/A for structure relaxation, and the energy convergence criterion was 10$^{-6}$ eV.A $\Gamma$ centered 9×9×9 Monkhorst-Pack grid was chosen for the electronic structural optimization. The phonon calculations were performed using Density Functional Perturbation Theory (DFPT) as implemented in VASP and the PHONOPY package \cite{gonze1997,togo2015}. The calculations were performed on the unit cell as we needed only the $q$ = 0 phonons, using a $7 \times 7 \times 7$ $k$-point grid. We mention that because of a complex spin structure, we refrained from using a Hubbard U correction in the simulation.

\section{Results and Discussion}\label{sec2}

\begin{figure}
\centering
\includegraphics[width= 0.7\linewidth]{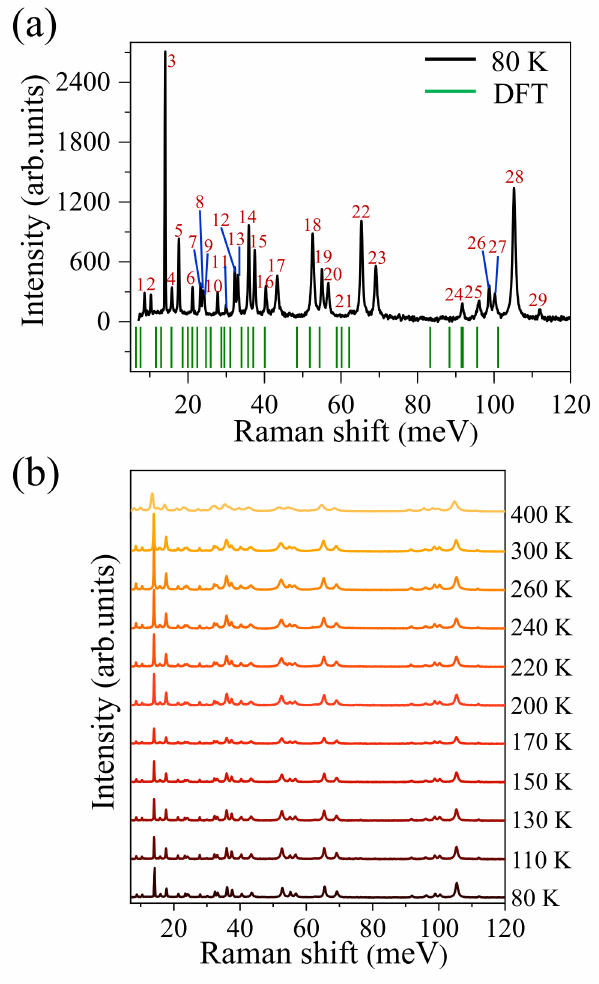}
\caption{(a) Raman spectrum of NCGO at 80 K over the spectral range 7.2 to 120 meV. All 29 Raman modes are marked. The green bars are the Raman peak positions at the $\Gamma$ point, obtained using non-spin-polarized DFT calculations. (b) Characteristic Raman spectra at various temperatures, marked in the right.}
\label{stack}
\end{figure}

Detailed X-ray and neutron diffraction studies revealing the crystal structure and phase purity of the compound are reported elsewhere \cite{Bera2022}. The Rietveld analyses of combined X-ray and neutron diffraction patterns reveal that the crystal structure of the compound belongs to the P-1 space group with triclinic symmetry.
The magnetic Cu ion (Cu$^{2+}$), and non-magnetic Na, Ge and O ions, are distributed at two, one, two, and six Wyckoff sites, respectively.

Figure \ref{stack}(a) plots the Raman spectrum obtained at 80 K over the spectral window between 7.2 and 120 meV. According to group theory, the irreducible representation for the Raman active modes at the $\Gamma$ point for this compound of space group P-1 is $\Gamma$=30A$\mathrm{_{1g}}$. We observe 29 Raman modes, marked as 1-29. Later in this article, we refer to them as P1, P2,...P29.    
From DFT calculations, we find 30 phonon modes, indicated by green bars in Fig. \ref{stack}(a), over the energy range of 0 to 120 meV. It is not surprising to find a significant mismatch between the simulated and experimental energies of the phonon modes, as the former involves a non-spin-polarized calculation for this complex correlated spin system. The phonon energy eigenvalues and eigenvectors associated with each mode are provided  in Supplementary Material  SM2 \cite{suppli}.  We would like to note that spin-polarized DFT calculations are computationally expensive, as they require a 168-atom supercell to accurately represent the existing linear spin-trimer chain of the antiferromagnet. The fundamental character of the phonon modes, including their irreducible representations and their atomic displacement patterns, is mainly governed by the crystal symmetry and bonding environments. Hence, the eigenvector analysis of non-spin polarized calculations remains qualitatively valid.

Raman spectra of the compound were recorded over a wide range of temperatures between 80 K and 400 K. Characteristic Raman spectra recorded at marked temperatures are shown in Fig. \ref{stack}(b).
We did not observe the appearance or disappearance of the Raman peaks, further confirming that the crystal structure of the one-dimensional trimer chain reported above remains stable over the entire temperature range of interest.

Quantum materials display the signatures of both their elementary and composite excitations in the broad background of their Raman spectral profile, as shown in earlier studies \cite{glamazda2016raman,Wulferding_2020,choi2021,Shainee2021,sokolik2022,sandilands2015scattering,Nasu2016,Pal2021,pal2022,PhysRevB.95.174429,takagi2019concept}. The light-matter interaction, involving the microscopic origin of such inelastic light scattering, is often entirely different from the magnons that occur in conventionally ordered magnetic materials. 
The dynamic response of phonons and underlying spin 
excitations in a quantum material at a finite temperature can be obtained from the imaginary part of the Raman susceptibility \cite{silveirinha2011,glamazda2016raman}
$\chi'' \propto $ $I(\omega)/ [1+n(\omega)]$ i.e., by normalizing the measured Raman spectral intensity $I(\omega)$ with the thermal Bose factor [1 + $n(\omega)$] = 1/(1-e$^{-h\omega/k_BT}$),  at a temperature $T$. While the broad continuum of $\chi ''$ carries the dynamical magnetic Raman response
of the spin excitations, the spectral profile of the phonon modes captures evidence of the coupling between phonon and quantum spin excitations in the system  \cite{pal2022}. Thus, the full spectral profile yields comprehensive information about the spin-lattice dynamics of the material.

In view of the contributions of multiple quasiparticle spin excitations to the Raman background, we fit the  Raman response of the recorded spectra at all temperatures over the full energy range between 7.2 meV and 120 meV using the relation \cite{yoon2000raman,singh2021fractional}
$\chi''(\omega,T)= \sum_{i=1}^{29}\frac{I^{i}_{\chi^{\prime\prime}}(T)}{2 \pi}\frac{\Gamma_{i}(T)}{(\omega-\omega_{i}(T))^{2}+ \left(\frac{\Gamma_{i}(T)}{2}\right)^{2}}+ Q(\omega,T)$. The first term models the observed 29 phonon modes using a Lorentzian function for each, with a Raman shift $\omega_{i}(T)$ , a width $\Gamma_{i}(T)$ and $I^{i}_{\chi^{\prime\prime}}(T)$ is the integral intensity at a temperature $T$. The second term, $Q(\omega,T) = F(T)\frac{\gamma(T)\omega}{\omega^{2}+\gamma^{2}(T)}$, describes the background intensity profile that originates from the semi-phenomenological model of quasi-elastic Raman scattering by different kinds of excitations, where $\gamma(T)$ measures the inverse lifetime of the excitations.  $F(T)$ is a constant for a temperature $T$. The best-fitted Raman responses of the spectrum recorded at 80 K, 120 K, 170 K, and 300 K are available in Supplementary  Fig. SF3 in SM3 of the SM \cite{suppli} . 
All spectra recorded at various temperatures using an excitation wavelength of 568 nm are also analyzed following the same procedure.

\begin{figure}[h]
\centering
\includegraphics[width= \linewidth]{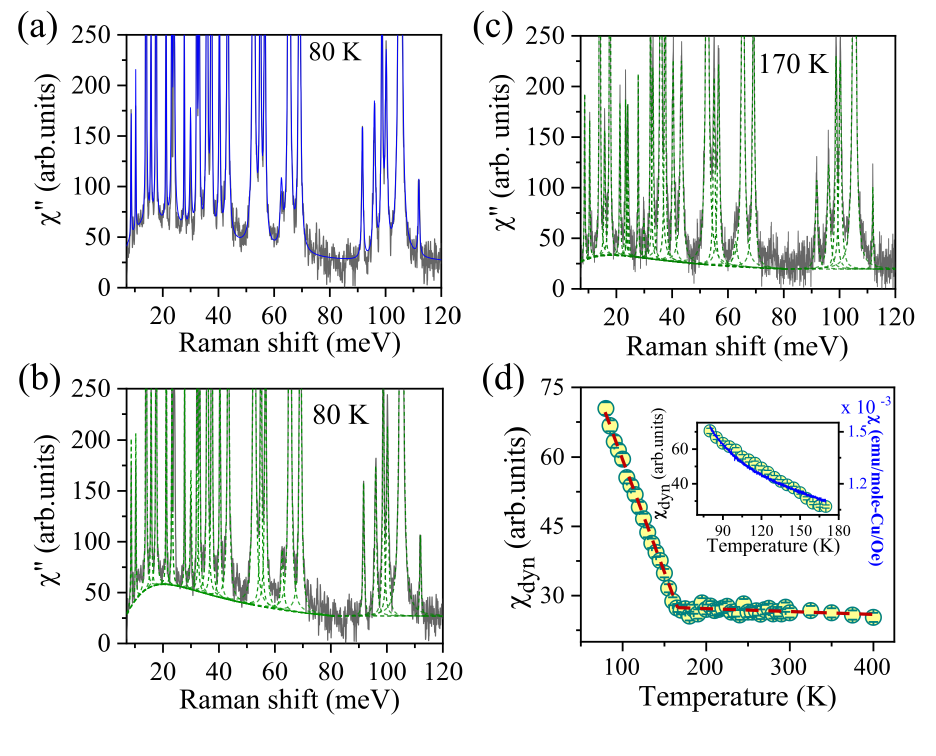}
\caption{
(a) Magnified view of the Raman response of the spectra recorded at 80 K. The recorded spectrum and the net fitted curve are shown by gray and blue, respectively.   (b) and (c) Deconvoluted components of the phonon modes and the background intensity profiles are shown by green dashed and green solid curves for the spectrum recorded at 80 K and 170 K, respectively. The value of $\gamma$ remains nearly unchanged at 18$\pm$1 meV between 80 K and 170
K. The recorded spectra are shown by gray curves. See text. (d) Dynamic spin susceptibility, $\chi_{dyn}$, as estimated from the integral intensity of the Raman background recorded at temperatures between 80 K and 400 K. The error bars (see SM5 \cite{suppli}) are within the size of the symbols. The dashed red line is the guide to the eyes.  The inset of the figure overlaps the estimated $\chi_{dyn}$ from Raman measurements (symbols), and the static magnetic susceptibility (blue curve).
}
\label{spectra_background}
\end{figure}

First, we focus our attention on broad Raman spectral background. The magnified view of the lower-intensity region of the Raman response at 80 K is shown in Fig.~\ref{spectra_background}(a). The contributions of phonon modes are highlighted by green-dashed curves in Fig. \ref{spectra_background}(b) for the spectrum recorded at 80 K. The same for 170 K is shown in Fig.~\ref{spectra_background}(c). In all panels, the gray curves represent the Raman response of the recorded spectra, while the blue curves depict the net fitted spectra.
 In panels (b) and (c), the fitted background profile (solid green curve) is traced along the crest of the phonon modes. A comparison of the broad Raman background in the spectra recorded at 80 K and 170 K using different excitation wavelengths is available in Supplementary  Material SM4 \cite{suppli}. The nearly identical background profiles without any shift with excitation wavelengths of 532 nm and 568 nm further confirm that their origin is more than trivial photoluminescence from the compound.
The cut-off energy of background intensity is found to be   $\sim 4J_1$, with $J_1=$20 meV \cite{Bera2022}.
Similar cutoffs in the Raman continuum have been reported for other quantum magnetic systems with different spin–spin interaction schemes.
The cutoff energy is determined by the finite bandwidth of the spin excitations, which is  governed by the dominant exchange interaction $J_1$, causing the continuum to terminate at an energy scale.
For example, in one-dimensional quantum spin-chain systems described by a $J_{1}-J_{2}$
 model—--such as CuGeO$_3$
and KCuF$_3$---
the broad Raman spectral background extends up tp 4$J_1$ where $J_1$ is the coupling strength between spins \cite{brenig1997,gnezdilov2012}. On the other hand,
Raman spectral background found earlier in  Kitaev spin-liquids \cite{sandilands2015scattering,Nasu2016,glamazda2016raman,PhysRevB.95.174429,takagi2019concept,Wulferding_2020,Pal2021,pal2022}, another class of quantum system, and has been explained by fractional excitations in terms of a pair of Majorana fermions up to the energy  $3J_K$, where $J_K$ is the Kitaev coupling strength \cite{Nasu2016,pal2022}. A theoretical understanding of the microscopic mechanism that gives rise to broad Raman background in the spin-1/2 trimer chain in NCGO is still lacking in the literature. 

\begin{figure*}
\centering
\includegraphics[width=0.7\linewidth]{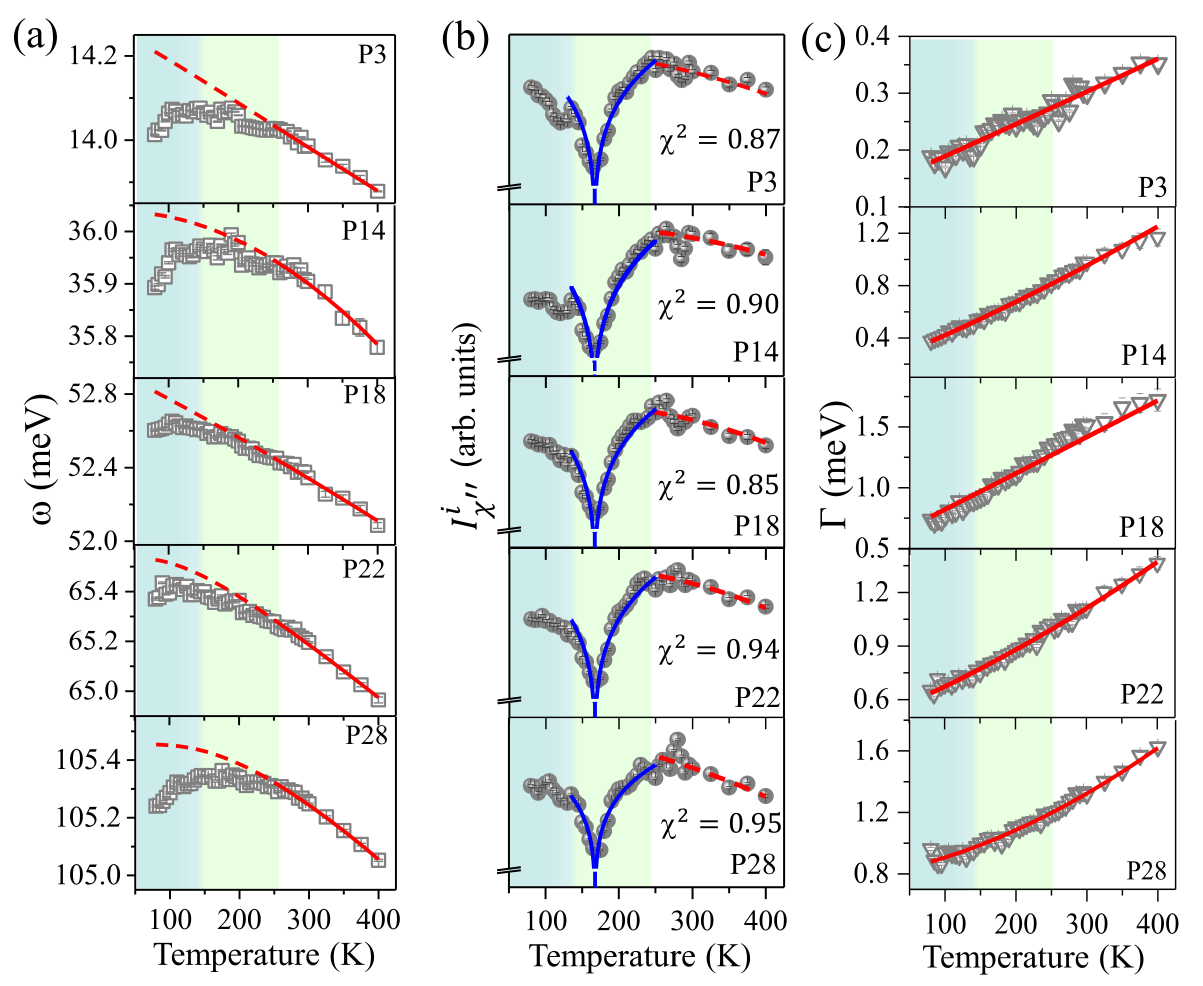}
\caption{(color online) 
Evolution of (a) Raman shift, (b)  integral intensity of $\chi^{\prime\prime}$, (c) FWHM of P3, P14, P18, P22, and P28 with temperature are shown by symbols. In panel (a), Raman shifts for 250 K$<T<$400 K are fitted  considering the anharmonicity in the change in vibrational potential with temperature (red solid curves) and then extrapolated down to 80 K (red dashed curves). In panel (b), the red dashed curves are guides to the eyes to the integral intensity of $\chi^{\prime\prime}$ above 250 K.  The intensity of Raman response in the pastel lime-shaded regions (for 130 K$\leq$T$\leq$250 K) follows a power law behavior  $I_{\chi^{\prime\prime}}^{i}\propto|T-T_{c}|^\beta$, with $\beta$= 0.24$\pm$0.02 and $T_c$=167$\pm1$ K (shown by blue curves). The intensity scales are shown with an axis break (double lines) to highlight  vicinity of the critical point $T_c$. The value of reduced $\chi^2$ in each panel, defined by
$\chi^2 = \frac{1}{(n-p)}\sum _{i=1}^n\frac{y_i-f(x_i)}{\sigma_i}$, describes the goodness of fit. Here, $n$= number of data points, $y_i$ is the actual observed value, $f(x_i)$ is the value obtained from the non-linear fit to the data points, and $\sigma_i$ is the standard deviation associated with that data. $p$ is the number of fitting parameters. In the present analysis, $n=25$, and $p=3$ for  $T_c$ and $\beta$. In panel (c), the FWHM over the entire temperature range increases monotonically with temperature. 
In all panels, the error bars to the data points are  the standard deviation of the spectral parameters, as obtained from the fitting procedure.
}
\label{Raman_Intensity_FWHM}
\end{figure*}

We estimate the dynamic spin susceptibility, as derived from the Kramers-Kronig relation, by integrating the Raman conductivity of the background spectral profile of the system as $\chi_{dyn}=\mathrm{lim}_{\omega\rightarrow {0}}\chi( k=0,\omega)\equiv\frac{2}{\pi}
\int{\frac{Q(\omega)}{\omega}d\omega}$ over 7.2 to 120 meV for each temperature and show in Fig. \ref{spectra_background}(d). The error bars for the data points are estimated from the error in the background profile, as obtained from the fitting procedure.

Also refer to the discussion in SM5. Due to thermal broadening, the reported \cite{Bera2022} range of $\leq$ 5 meV, 17 meV--22 meV, and 32 meV--37 meV, for spinons, doublons and quartons, respectively (measured at 3 K) spread over higher energy ranges, and fall within the energy scale of our Raman measurements, i.e.,  between 7.2 meV and 120 meV.  Thus, below $T_{s} \sim$ 170 K, the upturn of the plot relates to the existing spin dynamics of the system in the quantum magnetic phase, the trend of which matches the static magnetic susceptibility  (blue curve) \cite{Bera2022}, as shown in the inset. 
The correspondence between static and dynamic magnetic susceptibility in Fig. \ref{spectra_background} suggests that high-energy quasiparticle spin excitations in NCGO are quasi-static on the Raman time scale, 
and hence the dynamic response in Raman scattering effectively mirrors the equilibrium magnetic behavior.

We note that the background intensity is non-zero for $T \geq$ 170 K. We refer to the article \cite{yoshitake2016}, which discussed the dynamical spin fluctuations over the paramagnetic region in quantum systems following the Kitaev Hamiltonian. 
While in the quantum phase of these materials background Raman response arises due to  spin/magnetic excitations,
for the normal paramagnetic phase weaker structured background arises due  spin-fluctuation. The intensity of the latter remains nearly unchanged over the normal paramagnetic region in quantum spin liquid \cite{singh2021fractional, kumar2023,kumar2025} and Kitaev systems \cite{glamazda2016raman, Pal2021,pal2022}. The non-zero spectral background in the present study possibly reflects a similar effect. The static magnetic susceptibility of NCGO indicates that finite spin–spin correlations persist up to 
$\sim$ 300 K \cite{Bera2022}. Consistently, the temperature dependence of the quantum entanglement measure reported in Ref. \cite{Bera2022} suggests that these correlations extend up to 
$\sim$ 310 K, while the exchange interactions protect the system against thermal decoherence.
We see in Fig. \ref{spectra_background}(d) that although the $\chi_{dyn}$ remains constant for 170 K $\leq$ T $\leq$ 325 K; it exhibits a downward trend beyond 325 K (also refer to Fig. SF5(b)).

Next, we look into the characteristics of the sharp Raman modes in the spectral profile in Fig. \ref{stack}.  The evolution of the Raman shift with temperature of the five strong modes, P3 (at $\sim$ 14 meV), P14 (at $\sim$ 36 meV), P18 (at $\sim$ 53 meV), P22 (at $\sim$ 65 meV), and P28 (at $\sim$ 105 meV), is shown in Fig. \ref{Raman_Intensity_FWHM}(a).  The same is available for all other modes in Supplementary Fig. SF7(a) in Supplementary Material SM6 \cite{suppli}. The blue shift of the Raman modes with the lowering of temperature in the high temperature regime from 400 K can be attributed to the anharmonicity \cite{balkanski1983} in the lattice vibrations affecting the Raman shift in the normal paramagnetic phase of the system. However, a clear deviation from the monotonous increase in the Raman shift is observed below 250 K. The best fit to the data points between 250 K and 400 K, considering the anharmonic contribution to the lattice, is shown by the red solid curves in Fig. \ref{Raman_Intensity_FWHM}(a), assuming the negligible effect of the change in lattice volume with temperature.  The fitted curves, extrapolated down to 80 K, show the deviation of the data points below 250 K. In view of the existence of spin excitations over this low temperature range, the anomaly 
suggests a strong spin-lattice coupling below 250 K. A microscopic theoretical description of spin-phonon coupling in a one-dimensional spin-$1/2$ trimer system is not available in the literature. Consequently, a quantitative analysis of the deviations in the Raman shift aiming to extract the coupling strength between phonons and quasiparticle excitations in NCGO is beyond the scope of the present work.

We further report unusual anomalous behavior of the intensity of the Raman response, reflecting atypical underlying lattice dynamics of the trimer chain NCGO.  The evolution of the integral intensities of the Raman response with temperature for the five strongest Raman peaks: P3, P14, P18, P22, and P28, is shown in Fig. \ref{Raman_Intensity_FWHM}(b). The same for all relatively intense modes is available as Supplementary Fig. SF7(b) in Supplementary Material SM6 \cite{suppli}. With the lowering of the temperature from 400 K, the integral intensity of $\chi''$ increases down to 250 K due to the reduction in thermal decay channels for phonons in the system. In Fig. \ref{Raman_Intensity_FWHM}(b), the red dashed curves above  250 K are guides to the eye. An anomalous trend sets in below this temperature. The intensity of $\chi''$ drops and then turns up, exhibiting a dip at around 170 K. This trend is seen in all Raman peaks (see Supplementary Fig. SF7 (b) in Supplementary Material SM6 \cite{suppli}). The observed variation of the integral Raman response of all prominent phonon peaks over the temperature range $130\leq{T}\leq 250$ K could be fitted with a power law, $I_{\chi^{\prime\prime}}^{i}\propto|T-T_{c}|^\beta$, with the critical temperature as ($T_c$) and the critical exponent as ($\beta$) for a reasonable value of reduced $\chi^2$. The best fits to the data points are shown by blue curves (with the value of reduced $\chi^2$ in the inset) for peaks P3, P14, P18, P22, and P28 are available in Fig. \ref{Raman_Intensity_FWHM}(b), and for the rest of the Raman peaks in SF7(b). In statistical analysis, a reduced $\chi^2$ value close to 1 indicates that the model describes the data within the experimental uncertainties. Values much larger than 1 correspond to poor fits, while values much smaller than 1 can indicate overfitting to statistical noise \cite{data2002, livadiotis2025}. The reported $\chi^2$ values of 0.80–0.95 therefore indicate a satisfactory and statistically reasonable fit.
The values of $T_c$ and $\beta$ for individual modes, thus obtained, are available in Supplemental Table ST2 in SM6 of the SM \cite{suppli}.  
Averaging these values for all phonon modes in Supplemental Table ST2, the mean and standard deviation of $\beta$ is estimated as $\beta$= 0.24$\pm$0.02, with $T_c$= 167$\pm 1$ K.  It is to be recalled that the dynamic and static spin susceptibilities of the material overlap below $T_s$= 170 K (Fig. \ref{spectra_background}), and an earlier report \cite{Bera2022} demonstrated the existence of quasiparticle spin excitations (spinon, doublon, and quarton) over this low temperature region. 
The equivalence of $T_s$ and $T_c$ implies that the phonon renormalization observed across the crossover arises from coupling to slow and coherent spin dynamics rather than from fast and incoherent excitations.

To shed further light on the microscopic nature of the spin-lattice correlation within the critical scaling region, we examine the temperature variation of the Raman linewidth (full width at half maxima, FWHM). In general, magnetic excitations (magnon/spinon) introduce additional scattering pathways for phonons, leading to a loss in Raman intensity \cite { zhou2011,metavitsiadis2020}  
and concurrently increase the FWHM of the spectral line due to the opening of an additional decay channel. Interestingly, in the present system of interest, NCGO, the thermal variation of the FWHM of all modes exhibits a monotonic increase with temperature following the anharmonic model of vibrational potential over the entire temperature range of interest, as shown by the red solid curves in Fig. \ref{Raman_Intensity_FWHM}(c) and Supplementary Fig. SF7(c) in Supplementary Material SM6 \cite{suppli}. 
It suggests that the phonons do not find additional scattering center in the quantum phase. Thus, the observed nature of the phonon dynamics, as we observed in Fig. \ref{Raman_Intensity_FWHM}(b) for the integral intensity of the Raman response, is not the outcome of the existence of additional scattering channels due to spin excitations in the low-temperature region. We refer to Ref. \cite{pal2022},  which reports Raman fingerprints of fractionalized excitations in an irrdiate compound. In this article \cite{pal2022} authors demonstrated anomalous behavior of the intensity of the Raman response of the phonon modes, while the FWHM increased monotonically with temperature. In our case, too, for the 1D spin-$1/2$ trimer chain, we observe a similar effect, though in a different quantum system. 

Furthermore, we noticed that the spectral widths ($\Gamma$) of the low-energy peaks (P1--P10, which are below $\sim$ 28 meV) are extremely small in the low temperature range. At 80 K, for some of these low-energy phonon modes, it is just about in the range of our instrumental broadening, which is $\sim$ 0.17 meV [see supplementary material SM7 \cite{suppli,xu2004,letcher2007,daly2009}]. As 
$\Gamma =\frac{\hbar}{\tau
}$, where $\tau$ is the lifetime of the phonon modes, such long lifetimes of phonon modes are noteworthy for a material synthesized using the solid state reaction method, as followed by us. Keeping in mind that the energy scales of some of the spin-excitations in the system are within this range, the observation suggests that, at low temperature, the phonon-phonon anharmonic scattering channels become strongly suppressed, while phonons interact coherently with the quantum background.

\begin{figure*}
\includegraphics[width= 0.85\linewidth]{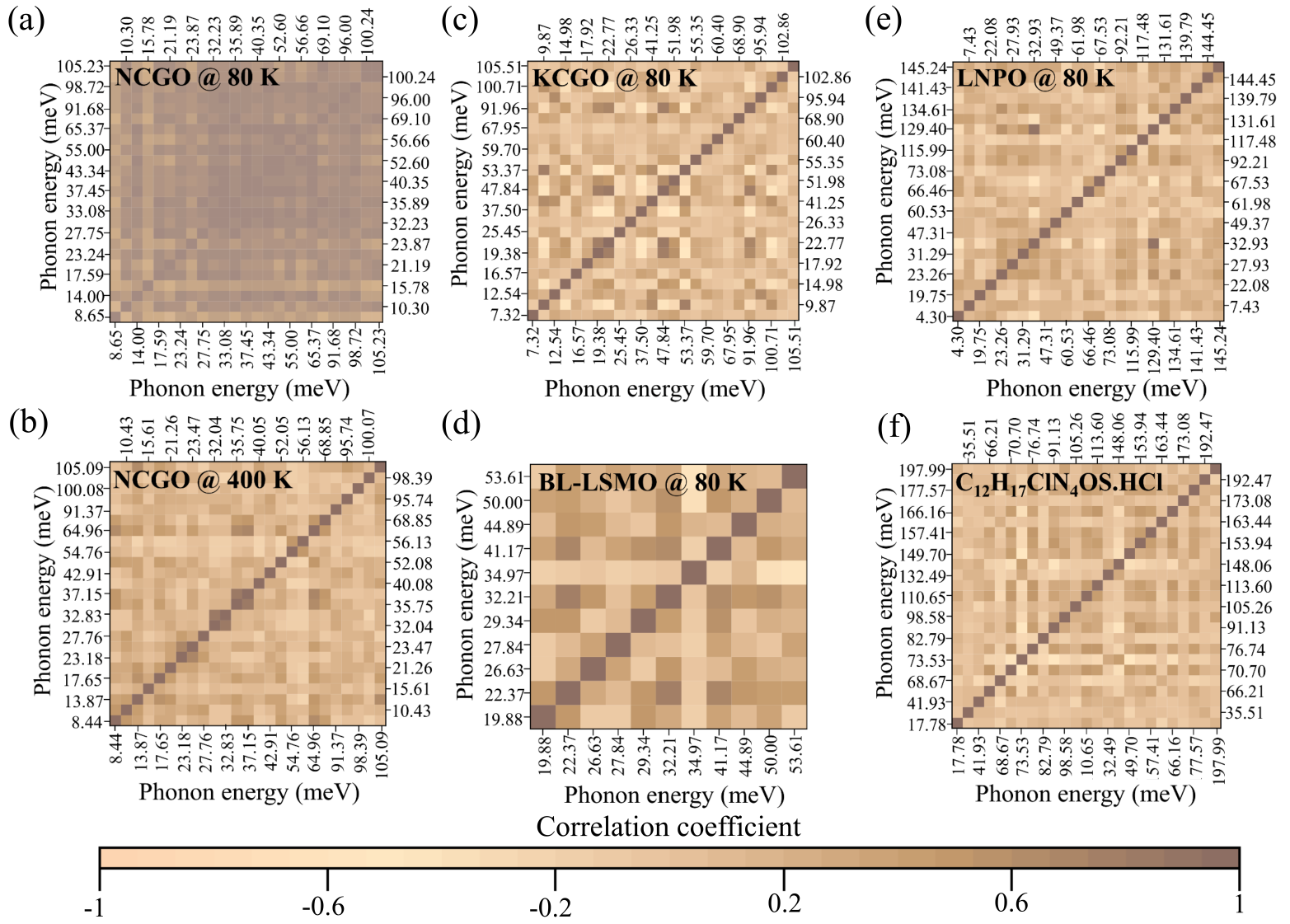}
\caption{ {(colour online) Correlation matrix for 200 Raman spectra recorded for (a) NCGO at 80 K, (b) NCGO at 400 K, (c) KCGO at 80 K (d) BL-LSMO at 80 K (e) LNPO at 80 K, and (f) an organic compound. For clarity, both top-bottom and left-right axes are used to refer to the median values (of a total of 200 measured values) of the Raman shift of alternate phonon modes.}}
\label{correlation}
\end{figure*}
Next, we highlight another distinct manifestation of spin-lattice coupling in NCGO, revealed through the statistical analysis of a large number of spectral data. 
Within a simple harmonic oscillator framework, the Hermitian nature of the Hamiltonian ensures that the phonon eigenstates are orthogonal. In conventional lattice thermal fluctuations, distinct symmetry representations of each phonon mode, the involvement of different atoms and their displacement patterns, and different anharmonic vibrational coefficients are expected to yield uncorrelated values of the energies of the phonon modes. 
Thus, there should be no statistical correlation among them. Interestingly, we observed a strong linear correlation among the phonon energy eigenvalues of NCGO at 80 K. 
 To reveal this correlation, we specifically performed a statistical analysis of the phonon eigenvalues obtained from Raman measurements at 80 K, at which spin excitations are present and couple with the phonon modes (as demonstrated above). For comparison, we also performed a similar analysis of the spectra recorded at 400 K, when the system is in the normal paramagnetic phase. 200 Raman spectra were recorded for each case and analyzed, following the same procedure mentioned for Fig. \ref{spectra_background}. 
 The linear correlation coefficient has been estimated for the $m\times n$ matrix, with $m$ = number of Raman modes ($m$ = 25 for NCGO; out of a total of 29 observed modes, 25 relatively strong modes, discussed earlier, are considered), and $n$ is the number of data sets, by following the Pearson correlation analysis. 
In this statistical analysis, the sample analog of the  correlation between two variables $X_i$ and $X_j$ is defined as 
\begin{equation}
r_{ij} = \frac {n\left(\displaystyle \sum_{l=1}^n X_{il} X_{jl}\right)-\left(\displaystyle \sum_{l=1}^n X_{il}\right)\left( \displaystyle \displaystyle \sum_{l=1}^nX_{jl}\right)}{\sqrt{\left[n \displaystyle \sum_{l=1}^n X_{il}^2- \left(\displaystyle \sum_{l=1}^n X_{il}^2\right)\right]\left[n \displaystyle \sum_{l=1}^n X_{jl}^2- \left(\displaystyle \sum X_{jl}^2\right)\right]}}.
\end{equation}
When $X_{il}$ and $X_{jl}$ are replaced by the corresponding realized (measured) values at a particular temperature. 
Here $n$ is the number of data points. 
Referring to Figs. \ref{correlation} (a) and (b), in which 
the pairwise correlation coefficient values of $r_{ij}$ for the 25 phonon modes of NCGO are mapped following the colour scale for their strength. The diagonal blocks correspond to the self-correlation of the phonon mode energies; hence, the value is 1. Interestingly, by looking into the non-diagonal terms, we observe a very high magnitude of correlation between the phonon energies in the case of NCGO at 80 K in (a). However, the correlation coefficient drops significantly at 400 K in (b).

To rule out possible measurement artifacts in establishing the strong correlation between the phonon frequencies at 80 K, several checks on our measured data were performed (see Supplementary Material  SM8 \cite{suppli,raj2021evaluating}).
Further, to explore whether the observed strong correlation between the phonon energies in Fig. \ref{correlation} is a trivial outcome of the intertwining of spin and lattice degrees of freedom, as observed in other magnetic systems, or due to the limited accuracy of measurements, we carried out a statistical analysis for three other systems: K$_2$Cu$_3$Ge$_4$O$_{12}$ (KCGO),  Li$_2$Ni$_3$P$_4$O$_{14}$ (LNPO), and pseudo-bilayer La$_{1.4}$Sr$_{1.6}$Mn$_2$O$_7$ (BL-LSMO), as all of these compounds exhibit a large number of Raman modes ($m$ = 25 for KCGO and LNPO, and $m$=11 for BL-LSMO), as shown in Supplementary Fig. SF12 in Supplementary Material SM9 \cite{suppli}. KCGO and LNPO are spin-$1/2$  and spin-$1$  trimer systems, respectively. While KCGO hosts a two-dimensional quantum spin-$1/2$ trimer mesh, which possesses spin excitations like NCGO,  LNPO is a two-dimensional trimerized collinear lattice antiferromagnet in which gapless quasielastic neutron scattering is observed at high temperatures ($T_{N}>$ 15 K). The dispersive and gapped magnon excitations exist below the transition temperature $T_N$ 
\cite{Chikara2025}. BL-LSMO belongs to the family of perovskite manganite magnetic systems, in which spin textures are reported \cite{yu2014}.  From neutron diffraction experiments, it is confirmed that BL-LSMO hosts spin waves up to 200 K  \cite{chatterji1999,chatterji1999spin}. For BL-LSMO, the evolution of spin-phonon coupling with temperature across multiple magnetic phases has been reported in the literature \cite{mekap2025}. For comparison, we also chose another non-magnetic organic compound (C$_{12}$H$_{17}$ClN$_4$OS.HCl) with $m$=25. For the statistical analysis, we recorded 200 Raman spectra of KCGO, LNPO, and BL-LSMO at 80 K. The same number of spectra of the organic compound were also recorded at room temperature. Each spectrum of all materials is deconvoluted using Lorentzian functions for the existing phonon modes to obtain phonon energies. The derived correlation matrices, obtained using the Pearson correlation analysis for 200 data sets of each system, are available in Fig. \ref{correlation}(c)-(f). 

\begin{table} [t]
\centering
\caption{
\label{extent} 
{Extent of correlation between the phonon energies as obtained from the statistical analysis.}
}
\begin{tabular}{cc}
\hline
  Compound & Extent of Correlation\\
\hline
NCGO  & 0.78 $\le \rho \le$ 0.87 (at 80 K) \\
  & 0.17 $\le \rho \le$ 0.27 (at 400 K) \\
\hline
KCGO & 0.19 $\le \rho \le$ 0.30 (at 80 K) \\
\hline
BL-LSMO & 0.20 $\le \rho \le$ 0.27 (at 80 K)  \\
\hline
LNPO & 0.16 $\le \rho \le$ 0.25 (at 80 K) \\
\hline
Organic compound & 0.15 $\le \rho \le$ 0.22 (Room temp.) \\
\hline
\end{tabular}
\end{table}

The following statistical analysis was performed to estimate the extent of correlation ($\rho$) between phonon energies in the systems mentioned above. In statistics, the Central Limit Theorem \cite{fischer2011} states that, under the influence of small and random perturbations (in the present case, the thermal fluctuations and the extent of reproducibility of measurements), the measured values of a variable follow a Gaussian distribution. We simulate  correlation matrices with off-diagonal  elements having a independently and identically distributed Gaussian distribution with a standard deviation of 0.05 and  a pre-assumed value of the correlation ($\rho$) ranging from $[0,1)$ changing with a fixed increment of $0.05$. The simulation is performed using a multivariate Gaussian distribution, and the matrix norms are then computed for the correlation of different values ($\rho$). The matrix norms of these simulated matrices are then approximated as a polynomial function of the correlation ($\rho$). Finally, a reverse mapping of the approximated polynomial is done to obtain one-point correlation as an inverse image of the norm of the correlation matrices obtained from the experimental data. Table 1 lists the estimated values of $\rho$, obtained from this statistical analysis for all studied compounds. Relevant plots and details are available in Supplementary Material SM10 \cite{suppli,rao1973linear}.  
The observed non-zero  values of $\rho$ in NCGO at 400 K, and in cases of KCGO, LNPO, BL-LSMO, and the organic compound, are possibly due to the limited size of the data set. On the other hand, the large value of $\rho$ for the spectral energy of NCGO at 80 K appears not to be an incidental correlation.

While recording a large number of spectra, one expects the value of the Raman shift for a particular peak to change at random, with a mean and standard deviation (say, $\omega$ and $\sigma$, respectively). The change in the value of the Raman shift for each mode in individual measurements is expected to be independent within the standard deviation. 
For the spectra recorded for all compounds, except NCGO  at 80 K, we find the same. Interestingly, in the case of spectra for NCGO at 80 K, the variation of Raman shift for all modes across all 200 data sets follows a linear association--if the Raman shift of a peak changes from $\omega\rightarrow\omega+\delta\omega$ in a measurement, the other peaks also change nearly by $+\delta\omega$. To further test whether such an observation is an artifact, we first estimated the mean peak position and standard deviation of all 25 Raman modes from the 200 measured data sets. By using the random number generation module in MATLAB, we generated 200 sets of values of Raman shift for each peak within the same value of the standard deviation. The extent of correlation of this artificially generated correlation matrix data dropped to 0.05 from $\sim$0.9 observed from the measured data. This indicates that, in the measured data, a factor exists that overcomes randomness and results in a linear correlation between the phonon energies. 

\section{Discussion}

Evidence of complex crossover between quantum and classical magnetic states has been established in the literature from renormalization group analyses and spin-fluctuation theories \cite{brando2016,moriya1973,millis1993,sachdev1999}.  In the present context, it is also relevant to note that real materials may exhibit crossover phenomena governed by multiple dynamical exponents due to finite disorder, as discussed in Ref.~\cite{brando2016}. Such complexity arises when quantum and thermal fluctuations play equally important roles, and conventional models of phonon dynamics are insufficient.

We infer the following:\\ (i) In Fig. \ref{spectra_background}, we have seen that below $T_s$=170 K, the Raman spectral background carries the signature of quasiparticle spin excitations in the system.  In contrast, earlier reports in the literature, based on INS measurements, demonstrated the presence of emergent spin excitations in the system (the measurement was carried out on the same sample) using dynamic structure factor analysis below $T$ = 250 K \cite{Bera2022}.
It is to be noted that in Raman scattering, while quasiparticles couple with photons and give rise to a broad continuum. However, the exact microscopic mechanism of Raman scattering by spin excitations in case of 1D spin-$1/2$ trimer is unknown. On the other hand,
INS probes the momentum-resolved dynamical spin structure factor $S(q,\omega)$. 
Additionally, the two techniques are sensitive to different dynamical time scales. The typical time scale probed in Raman scattering is of the order of 10$^{-12}$-10$^{-13}$ s, whereas INS
 accesses slower spin dynamics ($\sim 10^{-10}–10^{-12}$ s). Consequently, Raman scattering of phonons and INS may probe quasiparticle excitations with different sensitivities, depending on the underlying microscopic mechanisms. Thus, it is possible that the signature of coherent quasiparticle excitations could be observed only below 170 K in Raman measurements, while the same appear below 250 K in INS measurements.\\
(ii) From Fig. 4, we find that anomalies in the Raman spectral profile, evident in the phonon energy (panel (a)) and $I^{i}_{\chi^{\prime\prime}}$ (in (b)), set in at 
250 K. Notably, this is the same temperature below which INS measurements \cite{Bera2022} reveal the emergence of spin excitations in the system.
Thus, both INS and Raman scattering measurements indicate the deviation of the system from the normal paramagnetic phase below 250 K. However, the non-monotonic power law behavior of the intensity of Raman response of the phonon modes suggests a complex
interplay of local spin–lattice dynamics associated with the trimer-chain network over the temperature range 130 K $\leq T$ $\leq$ 250 K, with
$T_{c} \sim$ 170 K as the effective upper boundary of the critical scaling regime, delineating a crossover from the normal paramagnetic state to a quantum magnetic phase with emergent excitations.

As mentioned earlier, the present compound, NCGO, belongs to a special class of quantum materials that exhibit exotic quantum spin excitations. 
 We draw attention towards similar observations in other quantum magnets following various other types of spin-spin interactions.
A power law dependence of the intensity of the low energy continuum near the quantum critical point has been reported for the diluted Kitaev system $\alpha$-Ru$_{0.8}$Ir$_{0.2}$Cl$_3$ and has been attributed to the consequence of weak bond disorder in the system \cite{Do2020}. Furthermore, phonon renormalization and spectral weight redistribution near the quantum critical point often signal the presence of strong spin-lattice coupling and dynamic critical fluctuations \cite {Do2020}. Suppression of phononic heat transport in the quantum magnet BiCu$_2$PO$_6$, a spin-ladder compound, has been attributed to spin fluctuations near the quantum critical point as it attains low-energy elementary excitations \cite{jeon2016}. Critical scaling behavior of Raman spectral parameters as the system enters the magnetically ordered phase is observed for the coupled two-leg spin ladder Ba$_2$CuTeO$_6$ \cite{Glamazda2017}.
 The second-order phase transition  of the
S=1/2 square lattice antiferromagnet between the Néel state and the paramagnetic valence bond state is explained in terms of fractionalized degrees of freedom, emergent topological phenomena, and global conservation laws, rather than in terms of order parameters characterizing either state. In our studies on NCGO, the observed critical scaling of the integrated susceptibility of the phonon modes in Fig. \ref{Raman_Intensity_FWHM} does not necessarily signal quantum criticality in the system at the observed temperature $T_c$. 
However, the above-mentioned reports in the literature suggest that for systems  that typically lack conventional symmetry breaking, their phase transitions often fall outside the standard Landau-Ginzburg-Wilson framework \cite{Senthil2004,Grover2010}. 
It is reasonable to argue that the quantum criticality in the spin-$1/2$ trimer chain system, which carries fermionic quasiparticles, including spinons, doublons, and quartons, necessitates an alternative perspective. The non-availability of a theoretical understanding of the microscopic mechanism that drives the said emergent excitations in NCGO or its thermodynamic behavior leads us to surmise the following: rather
 than temperature, the onset of quantum fluctuations just below 250 K—when the system enters a region of non-trivial spin-spin correlation places the system among a broader class of low-dimensional quantum magnets where cooperative spin–lattice effects play a defining role in shaping the Raman response.

The above discussion complements the observed strong correlation between the phonon energies of the system. 
 It is important to note that, statistically, a non-zero correlation between two random quantities (for example, A and B) does not imply causation. Both A and B could depend on a ‘third party’ or latent variable, which can still lead to a correlation  of significant magnitude between them. 
The `third party,' which contributes to the high correlation between the phonon modes in the present NCGO compound at low temperatures, is quasiparticle spin excitations that strongly and uniquely couple with all phonon modes. 

\section{Summary}
Distinctive spin–lattice correlations and critical scaling behavior of the lattice dynamics establish NCGO as an exotic low-dimensional quantum magnet. Dynamic spin susceptibility, as obtained from the analysis of the Raman background, reveals that slow spin dynamics emerge below $T_s$= 170 K.  Strikingly atypical behavior of the phonon dynamics of the material is observed below 250 K.
 The temperature evolution of the integrated Raman susceptibility of the phonon modes demonstrates a crossover regime between the normal paramagnetic and quantum magnetic phases. Within the temperature range 130 K $\leq T \leq $ 250 K,  the same exhibits a critical scaling behavior characterized by a power-law dependence with a critical exponent $\beta$ = 0.24$\pm 0.02$ and critical temperature $T_c$ = 167$\pm1$ K. The proximity of $T_s$ and $T_c$
suggests that the observed phonon renormalization is driven by dynamic spin correlations. In addition, the relatively small values of FWHM of the low energy phonon modes further indicates that quasiparticle excitations interact coherently with phonons (hence,
increasing the lifetime of the phonon modes) rather than acting as a scattering channel, especially
at low temperatures. 
Furthermore, the compound exhibits an unusually strong correlation among phonon energy eigenvalues, as reflected in the high matrix-norm values of the correlation coefficients. This suggests a unique pathway of collective lattice response of the system influenced by the underlying spin correlations, providing deeper insight into the cooperative nature of spin-lattice interactions in low-dimensional quantum magnets.\\

\section*{Acknowledgments}
AR thanks Sudhansu Sekhar Mandal at IIT Kharagpur for the valuable discussion. AR acknowledges the financial support from DST, SERB, (CRG/2021/000718) GOI, and BRNS (58/14/24/2019-BRNS) GOI.  SMY thanks (58/14/24/2019-BRNS), GOI. SMY acknowledges the financial assistance from ANRF, Department of Science and Technology, Government of India, under the J C Bose fellowship program (JCB/2023/000014).

\section*{ Data availability}
The data analyzed in the current study are available from the corresponding author
on reasonable request.

\clearpage
\newpage
\onecolumngrid
\begin{center}
\textbf{\Large Supplementary Material for}\\[.3cm]
\textbf{\Large Phonon anomalies and  critical scaling in the spin-$1/2$ trimer chain Na$_2$Cu$_3$Ge$_4$O$_{12}$} \\[.4cm]


P. Srikanth Patnaik$^1$,  A. K. Bera$^{2,3}$, Srishti Bhardwaj$^4$,Tulika Maitra$^4$, Buddhananda  Banerjee$^5$, Anushree Roy$^1$, S.M. Yusuf$^{2,3,6}$ \\
\textit{$^1$Department  of  Physics, Indian Institute of Technology Kharagpur, Kharagpur 721302, India}

\textit{$^2$Solid State Physics Division, Bhabha Atomic Research Centre, Mumbai 400085, India}

\textit{$^3$Homi Bhabha National Institute,Anushaktinagar Mumbai 400094, India}

\textit{$^4$Department of Physics, Indian Institute of Technology Roorkee, Uttarakhand 247667, India.}

\textit{$^5$Department of Mathematics, Indian Institute of Technology Kharagpur, Kharagpur 721302, India}

\textit{$^6$UM-DAE Centre for Excellence in Basic Sciences, Vidyanagari, University of Mumbai, Mumbai 400098, India.}

\onecolumngrid

\end{center}

\setcounter{equation}{0}
\renewcommand{\theequation}{SE\arabic{equation}}
\setcounter{figure}{0}
\renewcommand{\thefigure}{SF\arabic{figure}}
\setcounter{section}{0}
\renewcommand{\thesection}{SM\arabic{section}}
\setcounter{table}{0}
\renewcommand{\thetable}{ST \arabic{table}}
\setcounter{page}{1}

\noindent This Supplementary Material consists of ten sections.\\
\noindent SM1: Optimization of the laser power on the sample.\\
\noindent SM2: Calculated energy eigenvalues and eigenvectors of different phonon modes of NCGO using first principles density functional theory. \\
\noindent SM3: Raman response of the spectra recorded at 80 K, 120 K, 170 K, and 300 K. \\
\noindent SM4: Comparison of the Raman response of the spectra recorded at  80 K and 170 K using 532 nm and 568 nm as the excitation wavelengths. \\
\noindent SM5: Estimation of dynamical spin susceptibility from the spectral background. \\
\noindent SM6: Variation of Raman shift, integral intensity, and full width at half maximum of the Raman response of the phonon modes with temperature. \\
\noindent SM7: Estimation of instrumental broadening and phonon lifetime of NCGO. \\
\noindent SM8: Tests prior to correlation analysis. \\
\noindent SM9: Raman spectra of KCGO, LNPO, BL-LSMO, and an organic compound C$_{12}$H$_{17}$ClN$_4$OS.HCl.\\
\noindent SM10: Correlation analysis.\\

\noindent \textbf{SM1: Optimization of the Laser Power on the Sample}\\

\begin{figure}[h]
\includegraphics[width=0.9\linewidth]{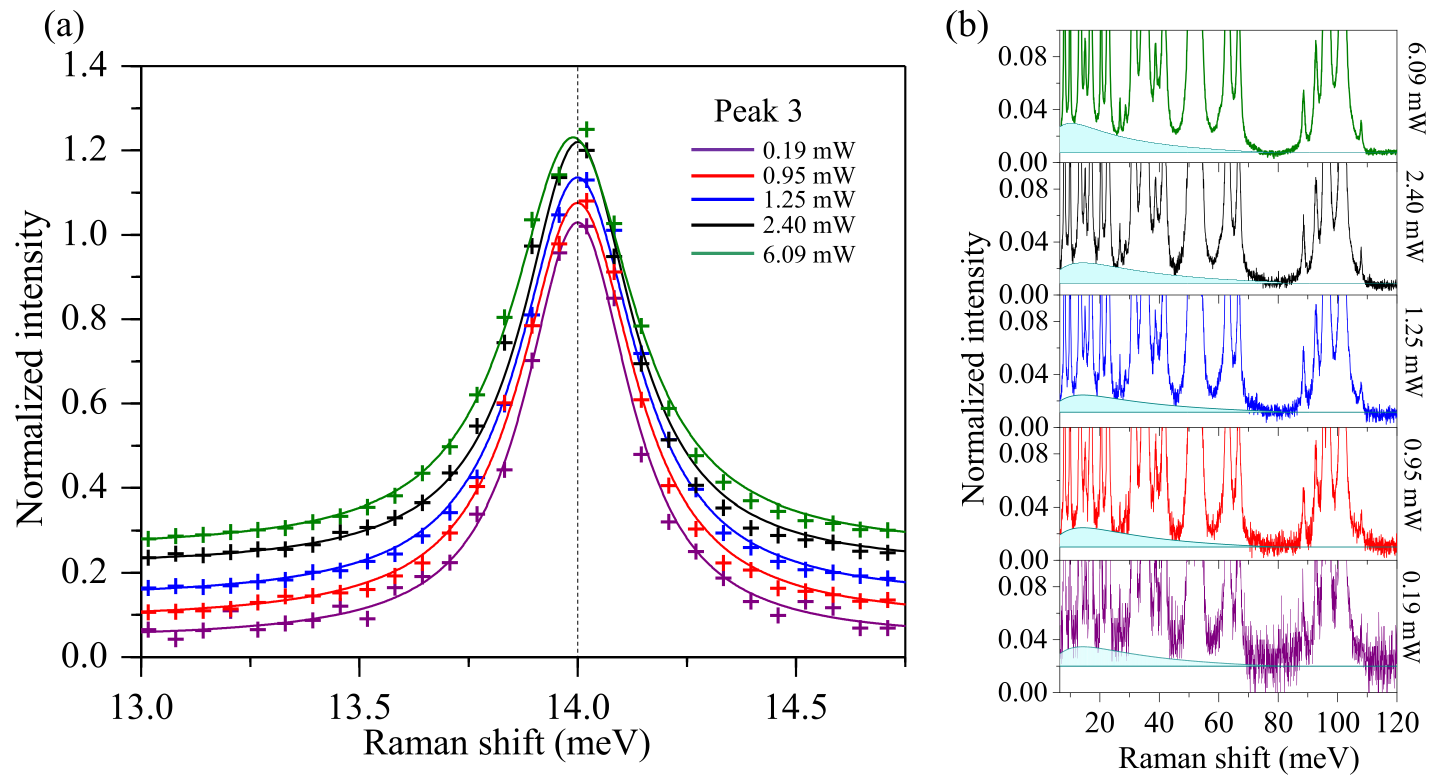}
\caption{ Comparison of Raman spectra of NCGO at 80 K with different laser powers for peak P3. (b) Comparison of the Raman background of NCGO at 80 K with different laser powers, mentioned on the right }
\label{Peak_shift}
\end{figure}

Before the reported measurements, we recorded spectra at different laser powers. Raman spectra of the most intense peak at 14.00 meV, recorded with different laser powers, are shown (data points are represented by + symbols, and the solid curves are the best fitted Lorentzian profile) in  \ref{Peak_shift}(a).  

We observe a slight red shift of  0.01 meV in the spectrum
recorded with 6.09 mW of incident laser power. The background was more prominently affected by the laser power, as shown in \ref{Peak_shift}(b). The background intensity substantially increased for laser power higher than 1.25 mW. As can be seen in  \ref{Peak_shift} (b), even with a laser power of 1.25 mW on the sample, the signal-to-noise ratio of the background is low For lower power, it dropped further. Hence, we chose a laser power of 1.25 mW to perform the experiments.

\newpage

\noindent \textbf{SM2: Calculated  Energy Eigenvalues and Eigenvectors of Different Phonon Modes of NCGO using First Principles Density Functional Theory}

\begin{table}[h!]
\caption{List of the experimental and calculated wavenumbers of all Raman active modes in NCGO. While the experimentally observed 29 modes are marked as Peak P1--Peak P29, the calculated modes are named as Mode 1--Mode 30.}
\centering
\scriptsize
\label{tab.dft} 
\begin{tabular}{cc}
\hline
\multicolumn{2}{c}{Experiment @ 80 K}\\
\hline
\hline
Peak & Peak Position\\
 & (meV)\\
\hline
 P1 & 8.65\\
\hline
P2 & 10.30\\
\hline
P3 & 14.00\\
\hline
P4 & 15.78\\
\hline
P5 & 17.59\\
\hline
P6 & 21.19\\
\hline
P7 & 23.24\\
\hline
P8 & 23.87\\
\hline
P9 & 24.19\\
\hline
P10 & 27.75\\
\hline
P11 & 29.94\\
\hline
P12 & 32.23\\
\hline
P13 & 33.08\\
\hline
P14 & 35.89\\
\hline
P15 & 37.45\\
\hline
\end{tabular}
\hspace{2em}
\begin{tabular}{cc}
\hline
\multicolumn{2}{c}{Experiment @ 80 K}\\
\hline
\hline
Peak & Peak position\\
 & (meV)\\
\hline

P16 & 40.35\\
\hline
P17 & 43.34\\
\hline
P18 & 52.60\\
\hline
P19 & 55.00\\
\hline
P20 & 56.66\\
\hline
P21 & 62.52\\
\hline
P22 & 65.37\\
\hline
P23 & 69.10\\
\hline
P24 & 91.68\\
\hline
P25 & 96.00\\
\hline
P26 & 98.72\\
\hline
P27 & 100.24\\
\hline
P28 & 105.23\\
\hline
P29 & 112.00\\
\hline
        &         \\
        
\end{tabular} 
\quad
\hspace{2em}
\begin{tabular}{cc}
\hline
\multicolumn{2}{c}{Calculated}\\
\hline
\hline
Mode & Mode position\\
Number & (meV)\\
\hline
Mode 1 & 6.39\\
\hline
Mode 2 & 7.60\\
\hline
Mode 3 & 11.60\\
\hline
Mode 4 & 12.95\\
\hline
Mode 5 & 15.65\\
\hline
Mode 6 & 18.59\\
\hline
Mode 7 & 19.95\\
\hline
Mode 8 & 21.11\\
\hline
Mode 9 & 22.41\\
\hline
Mode 10 & 24.70\\
\hline
Mode 11 & 25.93\\
\hline
Mode 12 & 28.72\\
\hline
Mode 13 & 29.50\\
\hline
Mode 14 & 31.04\\
\hline
Mode 15 & 33.99\\
\hline
\end{tabular}
\hspace{2em}
\begin{tabular}{cc}
\hline
\multicolumn{2}{c}{Calculated}\\
\hline
\hline
Mode & Mode position\\
Number & (meV)\\
\hline
Mode 16 & 35.70\\
\hline
Mode 17 & 37.06\\
\hline
Mode 18 & 40.08\\
\hline
Mode 19 & 48.48\\
\hline
Mode 20 & 51.86\\
\hline
Mode 21 & 54.41\\
\hline
Mode 22 & 58.89\\
\hline
Mode 23 & 60.16\\
\hline
Mode24 & 62.15\\
\hline
Mode 25 & 83.29\\
\hline
Mode 26 & 88.35\\
\hline
Mode 27 & 91.50\\
\hline
Mode 28 & 91.86\\
\hline
Mode 29 & 95.60\\
\hline
Mode 30 & 101.07\\
\hline
\label{phonon_eigen}
\end{tabular}
\end{table}

\vspace{3cm}
\begin{figure}[t!]
\centering
\includegraphics[width=\linewidth]{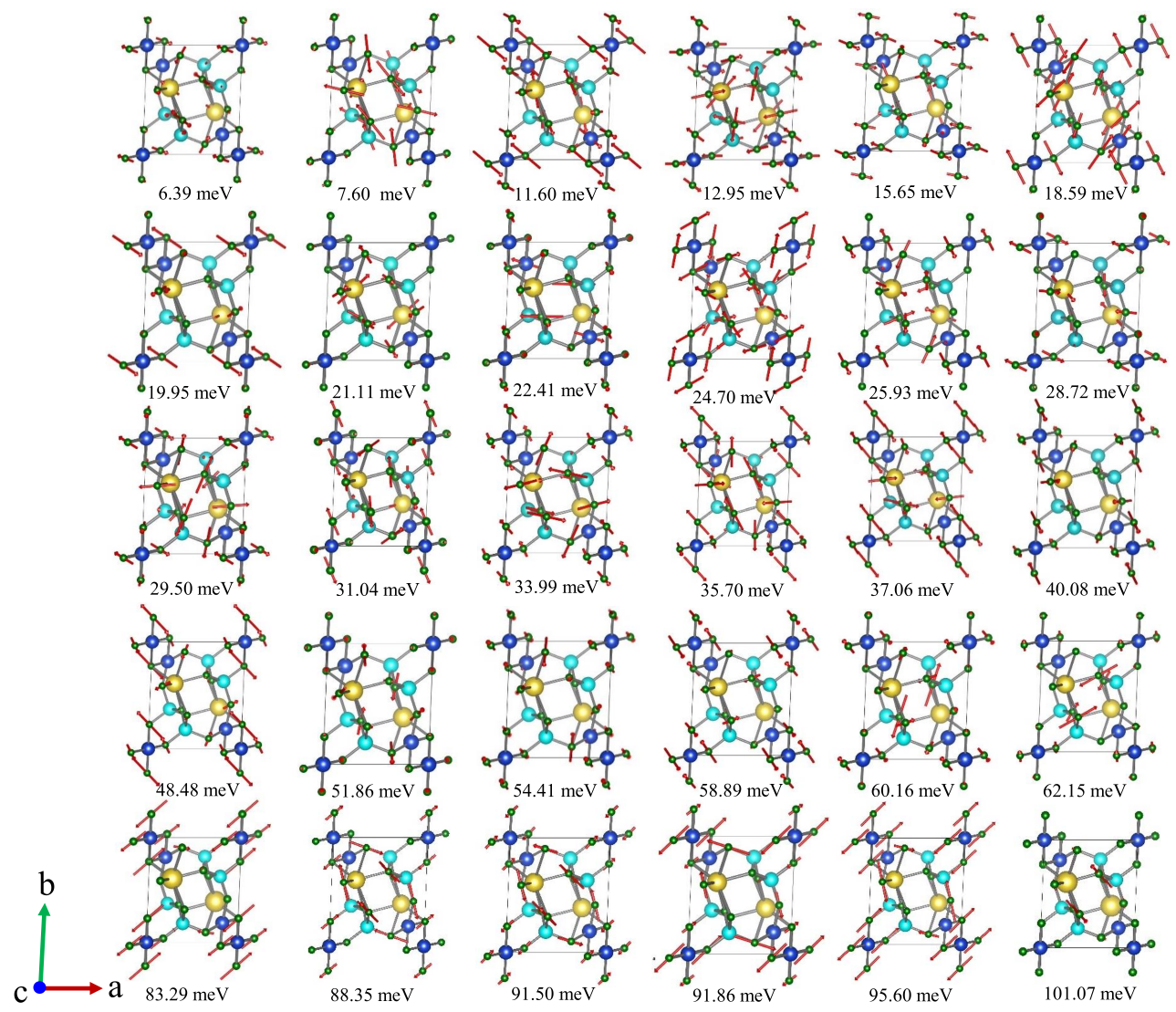} 
\caption{Atomic vibrations for mode numbers 1-30 of \ref{phonon_eigen}. The yellow, blue, cyan, and red color symbols represent sodium, copper, germanium, and oxygen, respectively. The red arrow represents the eigenvectors corresponding to each eigenvalue tabulated in \ref{phonon_eigen}}
\end{figure} 

\clearpage
\noindent \textbf{SM3: Raman Response of the Spectra Recorded at 80 K, 120 K, 170 K, and 300 K}

\begin{figure}[h]
\centering
\includegraphics[width= 0.9\linewidth]{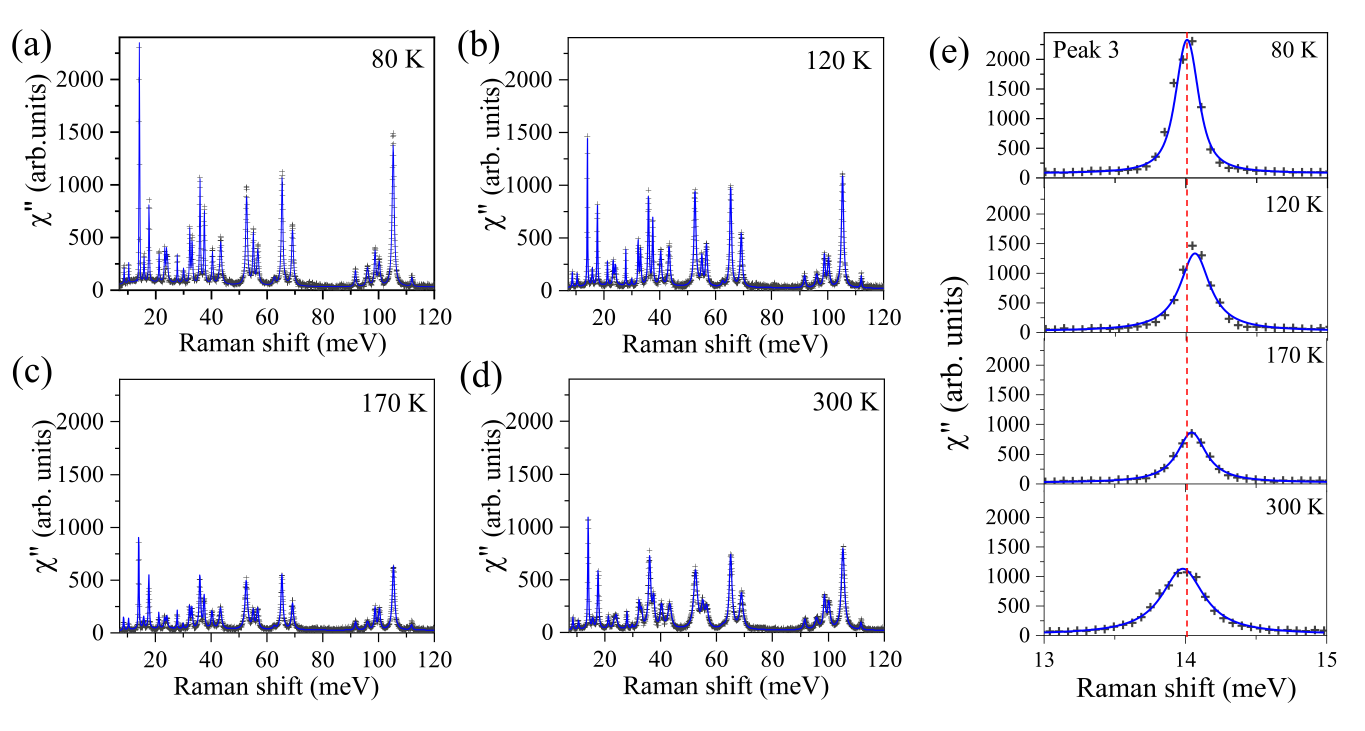}
\caption{Raman response of the spectra recorded at (a) 80 K, (b) 120 K,(c) 170 K, and (d) 300 K are shown by + symbols. (e) Evolution of peak P3 with different temperature.  The best fitted spectra, following the procedure mentioned in the main text, are shown by blue curves.}
\label{response}
\end{figure}

\newpage
\noindent \textbf{SM4: Comparison of the Raman Response of the Spectra Recorded at 80 K and 170 K using 532 nm and 568 nm as the
Excitation Wavelengths.}\\

\begin{figure}[h]
\centering
\includegraphics[width= 0.85\linewidth]{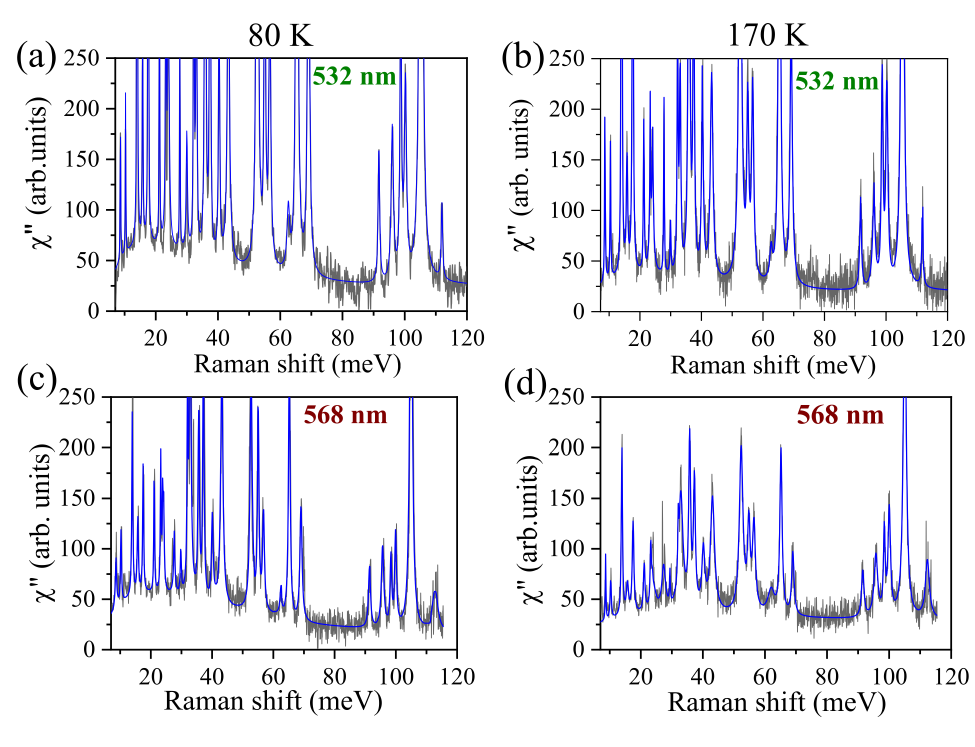}
\caption{ Magnified view of the Raman response of the spectra recorded at (a) 80 K and (b) 170 K using 532 nm as the excitation wavelength (the same as available in the main text); (c) 80 K and (d) 170 K using 568 nm as the excitation wavelength. In all the panels, the gray and blue curves are the raw data and the net fitted spectrum, respectively. The same fitting procedure was followed as in Fig. 3 of the main text. }
\label{568nm}
\end{figure}
 
\vspace{0.1cm}

In Raman scattering, the intensity of a phonon mode drops as $1/\lambda^4$, where $\lambda$ is the excitation wavelength. Thus, the intensities of phonon modes for the spectra recorded with 568 nm ((c) and (d)) are significantly less than those in (a) and (b), respectively. The broad background in a Raman spectrum may arise due to the luminescence of the compound. However, in such a case, the peak of the broad feature is expected to shift with the excitation wavelength.  The nearly identical spectral background indicates that its origin is not trivial photoluminescence from the material.

\vspace{1.5cm}

\noindent \textbf{SM5: Estimation of Dynamical Spin Susceptibility from the Spectral Background}\\

\begin{figure}[h]
\centering
\includegraphics[width= \linewidth]{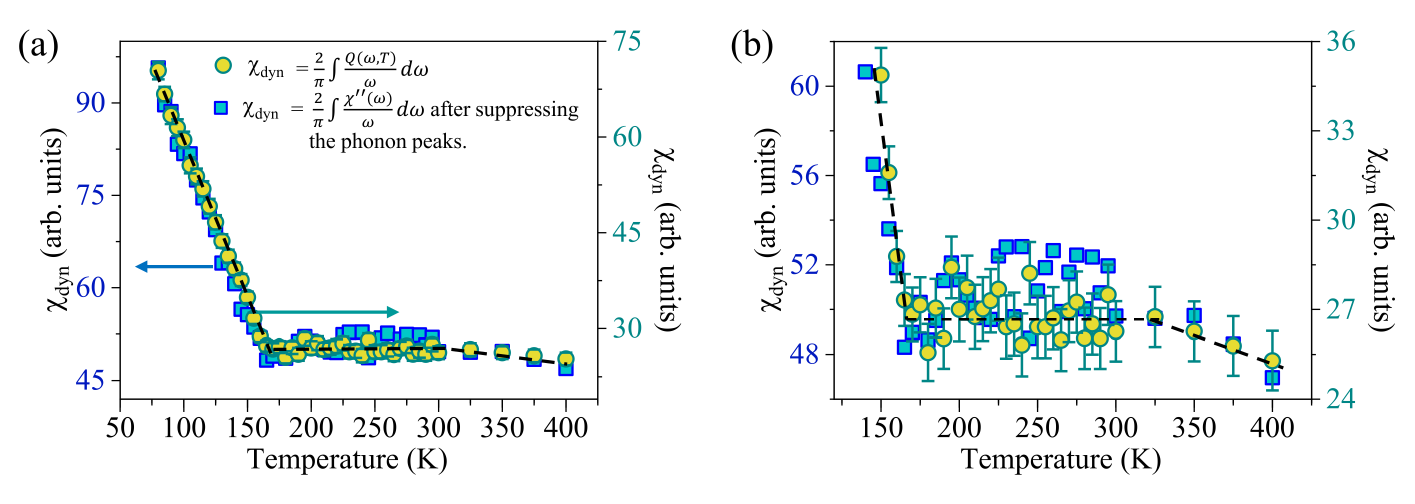}
\caption{(a) Comparison of $\chi_{dyn}$ obtained from fitting the background (in yellow circles) and by suppressing the phonon peaks (by blue squares). (b) Magnified view of $\chi_{dyn}$ to highlight its drop in arbitrary units above 325 K.} 
\label{dynamical}
\end{figure}

In addition to the method described in the main text, we have also estimated $\chi_{dyn}$ using the following procedure. We estimate the dynamic spin susceptibility by integrating the Raman conductivity ($\chi^{\prime\prime}/\omega$) as $\chi_{dyn}=\mathrm{lim}_{\omega\rightarrow {0}}\chi( k=0,\omega)\equiv\frac{2}{\pi}
\int{\frac{\chi^{\prime\prime}(\omega)}{\omega}d\omega}$ over 7.2 to 120 meV of the background spectral profile (after suppressing the phonon peaks [26,40] ), and show it with blue square symbols in Fig. \ref{dynamical}(a). The same obtained following the procedure described in the main text is shown by yellow symbols for comparison. The estimated $\chi_{dyn}$ following both procedures exhibits a similar trend. It turns up below 170 K. Refer to the magnified view of the plots, as obtained from both procedures, in Fig. \ref{dynamical}(b). $\chi^{\prime\prime}$ remains nearly constant over $170~K <T< 325~K$. Above 325 K the data points show a downward trend.  Because of the drop in recorded Raman intensity at high temperatures, the relatively large error bars to the data points could not be avoided.

\vspace{0.5cm}

\begin{figure}[h]
\includegraphics[width=0.45\linewidth]{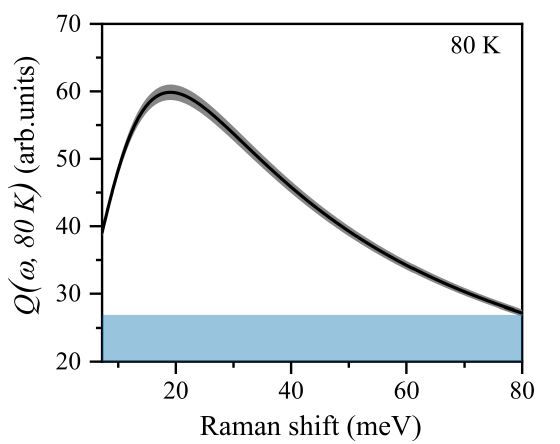}
\caption{Extracted spectral background component from the fitted parameters $Q(\omega$,80 K). The thickness of the curve represents the error bar in estimating it, following the spectral parameters obtained from the fitting procedure. The shaded background is not accounted for in the analysis.}
\label{fitting}
\end{figure}

\noindent Sensitivity test: For the procedure described in the main text, we could carry out the sensitivity test. The Raman background profile $Q(\omega)$ at 80 K, as extracted from the fitting procedure, is shown in Fig. \ref{fitting}. The thickness of the $Q(\omega)$ curve represents the error bar in estimating it. The estimated $\chi_{dyn}$= 70$\pm$1 arb. units.

\clearpage
\noindent \textbf{SM6: Variation of Raman Shift, Integral Intensity, and Full Width at Half Maximum of the Raman Response of the Phonon Modes with Temperature}
\begin{figure}[h]
\includegraphics[width= 1.0\linewidth]{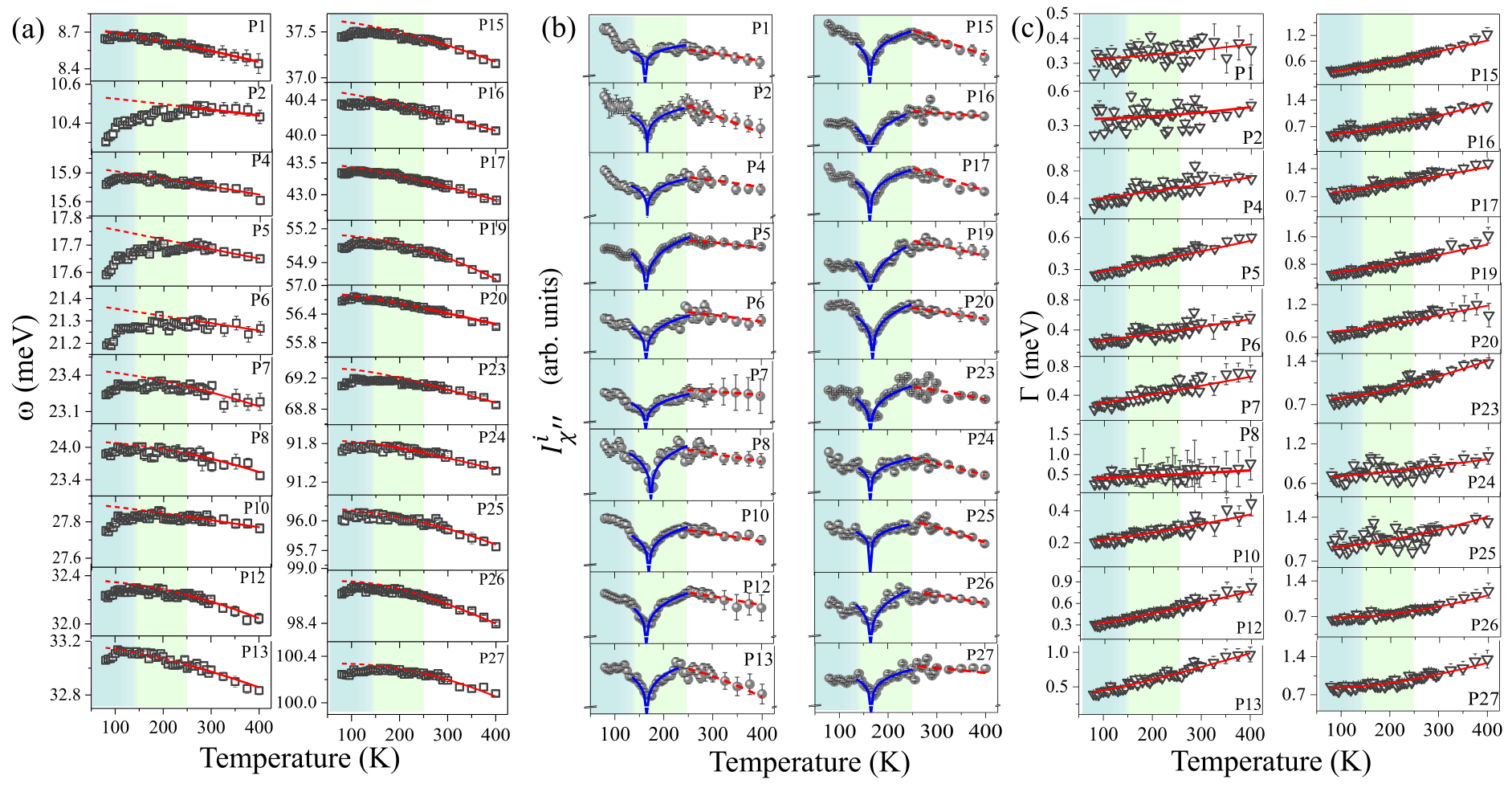}
\caption{Evolution of (a) Raman shift, (b) the integral intensity of the Raman response, (c) the spectral width of the observed relatively intense phonon modes with temperature. The same for P3, P14, P18, P22, and P28 are available in the main text.  For (a), the red solid curves are the best fits to the Raman shifts between 400 K and 250 K, considering the anharmonic contribution of the vibrational potential to the Raman shift [58]. The extrapolated curves down to 80 K are shown as dashed curves. 
In panels (b), the integral intensity of the Raman response in the pastel lime-shaded regions follows a power law behavior  $I_{\chi^{\prime\prime}}^{i}\propto|T-T_{c}|^\beta$ (shown by blue curves). The red dashed curves are guides to the eye for $T>250 K$. In (c), the solid red curves represent the best fit to the data points, showing the evolution of Raman line widths with temperature, taking into account the effect of anharmonicity, as in (a).   \\The peak positions, intensity, and width of the modes P9, P11, P21, and P29 are not shown, as P9 appears as a shoulder peak to P8 (and hence, the error bars to the  data points are high), while the rest exhibit weak intensities.
In all panels, the error bars on the data points, as obtained from the fitting procedure, represent the standard deviation of the spectral parameters and are within the size of the symbols.}
\label{Intensity:}
\end{figure}
\vspace{2cm}

\begin{table}[h]
\caption{List of critical temperature, critical exponents and Reduced $\chi^2$ obtained from fitting the integral intensity of Raman response of the intense modes over the temperature range 130 K to 250 K with $I_{\chi^{\prime\prime}}^{i}\propto|T-T_{c}|^\beta$}
\centering
\label{critical} 
\begin{tabular}{cccc}
\hline
Peak & $T_C$ (K) & $\beta$ & $\chi^2$\\
\hline
P1 & 167.79 $\pm$ 1.09  & 0.20 $\pm$0.02 & 0.82\\
\hline
P2 & 167.82 $\pm$ 1.08  & 0.24 $\pm$ 0.01  & 0.86 \\
\hline
P3 &166.79 $\pm$ 0.59  & 0.24 $\pm$ 0.02 & 0.87 \\
\hline
P4 & 167.29 $\pm$ 1.12  & 0.25 $\pm$ 0.01   &0.80 \\
\hline
P5 &166.91 $\pm $0.58  & 0.25 $\pm$0.02  & 0.92 \\
\hline
P6 & 167.29 $\pm$ 0.89   & 0.26 $\pm$ 0.02   & 0.80\\
\hline
P7 & 166.17 $\pm$ 0.89  & 0.24 $\pm$ 0.01  & 0.81  \\
\hline
P8 & 167.79 $\pm$ 0.65   & 0.23 $\pm$ 0.03  & 0.86 \\
\hline
P10 & 167.25 $\pm$ 1.09  & 0.22 $\pm$ 0.03   & 0.81 \\
\hline
P12 & 167.34 $\pm$ 0.64 & 0.24 $\pm$ 0.01  & 0.83 \\
\hline
P13 & 167.57 $\pm$ 0.56  & 0.26 $\pm$ 0.02  & 0.86 \\
\hline
P14 & 166.77 $\pm$ 0.64  & 0.28 $\pm$ 0.02  & 0.90 \\
\hline
P15 & 167.32 $\pm$ 0.64 & 0.23 $\pm 0.01$  & 0.91\\
\hline
\end{tabular}
\hspace{0.5cm}
\begin{tabular}{cccc}
\hline
Peak & $T_C$ (K) & $\beta$ & $\chi^2$\\
\hline

P16 & 166.66 $\pm$ 1.08 & 0.23 $\pm 0.04$  & 0.81 \\
\hline
P17 & 167.16 $\pm$ 0.69  & 0.25 $\pm$ 0.02 & 0.85 \\
\hline
P18 & 166.95 $\pm$ 0.57  & 0.23 $\pm$ 0.01 & 0.85\\
\hline
P19 & 166.99 $\pm$ 0.80 & 0.24 $\pm$ 0.02 & 0.86 \\
\hline
P20 & 166.17$\pm$ 0.69  & 0.27 $\pm$ 0.04 & 0.94 \\
\hline
P22 &167.08 $\pm$ 0.60  & 0.23 $\pm$ 0.01 & 0.94\\
\hline
P23 & 167.62 $\pm$ 0.82  & 0.25 $\pm$ 0.02  & 0.85\\
\hline
P24 & 166.98 $\pm$ 1.16  & 0.23 $\pm$ 0.01 & 0.82 \\
\hline
P25 & 167.31 $\pm$ 1.36 & 0.24 $\pm$ 0.01 & 0.85 \\
\hline
P26 & 167.58 $\pm$ 0.95  & 0.24 $\pm$ 0.02  & 0.83 \\
\hline
P27 & 168.41 $\pm$ 1.06 & 0.23 $\pm$ 0.01  & 0.85 \\
\hline
P28 & 167.53 $\pm$ 0.51 & 0.29 $\pm$ 0.02 & 0.95\\
\hline
& & &
\end{tabular}
\end{table}

\newpage
\noindent \textbf{SM7: Estimation of Instrumental Broadening and Phonon Lifetime of NCGO} \\
\begin{figure}[h]
\includegraphics[width=0.35\linewidth]{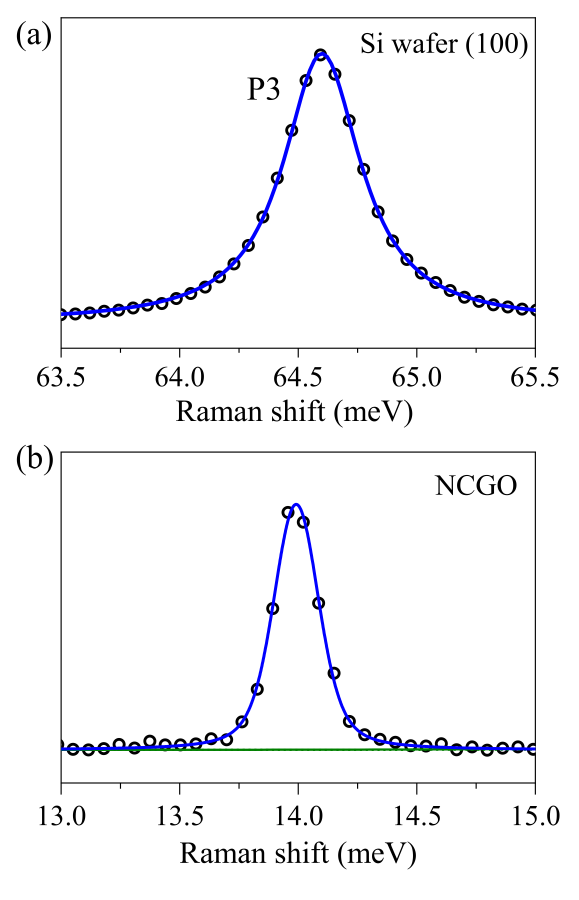}
\caption{Raman spectra of (a) Silicon wafer (100) and (b) NCGO recorded at 80 K showing peak 3. The recorded data and net-fitted spectra are shown as black circles and blue curves, respectively.}
\label{width}
\end{figure}

The full width at half maximum ($\Gamma$) and the lifetime ($\tau$) of the phonon mode are related by \\
\begin{equation}
\Gamma=\frac{\hbar}{\tau }
\label{lifetime}
\end{equation} 

To estimate the instrumental broadening, we fit the Raman spectrum of the calibration silicon wafer, as shown in Fig. \ref{width} (a), using a Voigt function.  The spectral width ($\Gamma_{ins}$) of the Gaussian component of the function is due to the instrumental broadening, and the width ($\Gamma_{phn}$) of the Lorentzian component is the intrinsic FWMH of the phonon mode $\Gamma_{phn}$. The value of  $\tau_{phn}$ of silicon $\sim$ 2 ps [63--65]. Using Eqn. \ref{lifetime}, the intrinsic FWHM of silicon $\Gamma_{phn}$ = 0.32 meV. By fixing this value as the Lorentzian width in the Voigt fitting of the silicon spectrum, we obtain the instrumental broadening $\Gamma_{ins}$ = 0.17 meV.

Thereafter, we fitted the peak P3 of NCGO (Fig. \ref{width} b) with a Voigt function, keeping instrumental broadening fixed at $\Gamma_{ins}$ =0.17 meV. From the fitting, we obtained the intrinsic phonon width of NCGO  $\Gamma_{phn}$ =0.09 meV, which corresponds to a phonon lifetime ($\tau \sim $ 7.5 ps). Such a long lifetime for a phonon mode is noteworthy for a strongly correlated polycrystalline material at 80 K. 

\vspace{0.5cm}

\noindent \textbf{SM8: Tests Prior to Correlation Analysis} \\

\begin{figure}[h]
\includegraphics[width=0.4\linewidth]{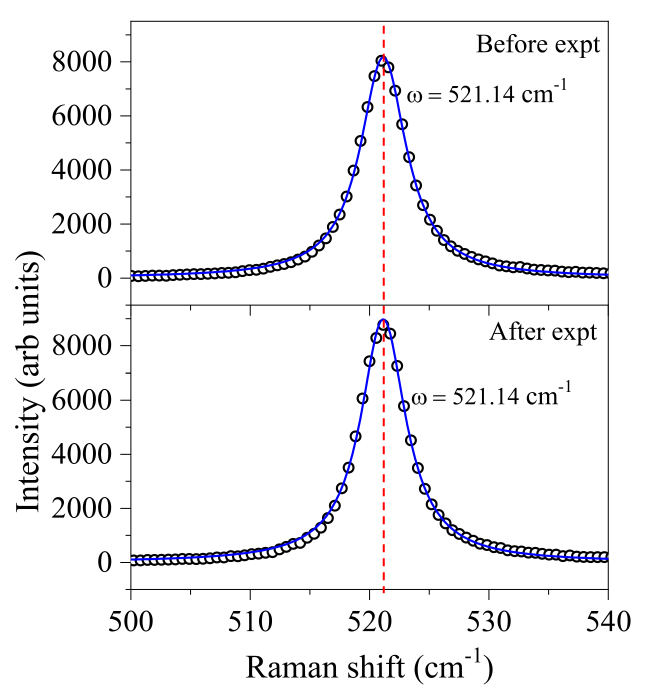}
\caption{Raman spectra of silicon (100) calibration wafer recorded at the start and at the end of 200 data records of NCGO. The recorded data and net-fitted spectra are shown as black circles and blue curves, respectively. The red dashed line marks the peak positions of two spectra, establishing the stability of measurements.  }
\label{sillicon}
\end{figure}

\noindent\textbf{(a) Spectrometer calibration test:} \\
  For solid samples, to calibrate the spectrometer, we use a Si(100) calibration standard sample. We tested the calibration stability of the spectrometer/laser/detector by recording Si spectra at the beginning and end of each experiment. Fig. \ref{sillicon} below compares the Si spectra recorded at the start and at the end of the 200 data records of NCGO for which a correlation analysis has been performed. The nearly identical spectral profile of Si confirms the stability of the spectrometer calibration during measurements.

\noindent\textbf{(b) Sample-temperature stability test:} \\

\begin{figure}[h]
\includegraphics[width=0.3\linewidth]{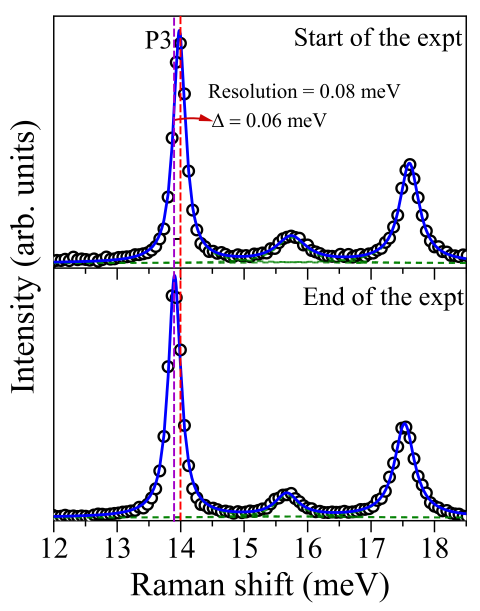}
\caption{First and last recorded Raman spectra of NCGO at 80 K showing peaks P3, P4, and P5. The recorded data, net fitted spectra, and deconvoluted components are shown as black circles, blue curves, and green dashed curves, respectively. The red and violet dashed lines marks peak position of P3 with a marginal change in Raman shift ($\Delta$ = 0.06 meV).}
\label{ncgo}
\end{figure}

The first and the last Raman spectra recorded while measuring 200 spectra of NCGO are compared in Fig. \ref{ncgo}.  A systematic rise in sample temperature is expected to red shift the peaks. The nearly identical spectral profile confirms the stability of the sample temperature. 
After recording each spectrum for 30 sec (integration time) at each temperature, we waited for 2 mins (laser cut-off) and recorded the next spectrum. \\

 \noindent\textbf{(c) Tests for global shifts:} \\

 \begin{figure}[h]
\includegraphics[width=0.7\linewidth]{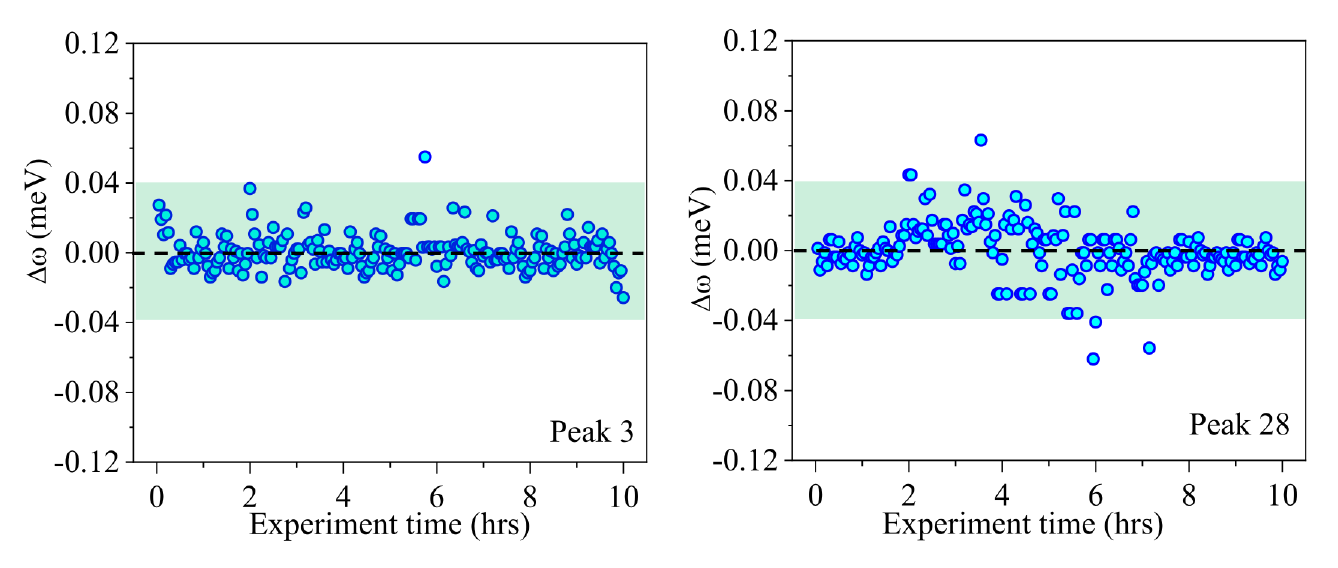}
\caption{Deviation of peak position of (a) peak P3, and (b) peak P28, over time when 200 Raman spectra were recorded at 80 K. The black dashed lines in both panels is the zero deviation line. The green shaded region is within instrumental resolution.}
\label{global}
\end{figure} 
Few reports in literature discusses the global shift in peak positions due to changes in physical parameters such as temperature, pressure, and humidity, or to instrumental artifacts such as spectrometer stability or a change in laser wavenumber [66]. 
 To examine if any of these effects are present during the recording of 200 spectra of  NCGO for correlation analysis at 80 K, Fig. \ref{global} below plots the deviations in the peak positions ($\Delta \omega$) of peaks P3, and P28 as a function of experimental time. Here $\Delta \omega$ = $\omega_t^n$ - $\omega_{avg}$, $\omega_t^n$ is the peak position of $n^{th}$ peak, recorded at a particular time ($t$) and $\omega_{avg}$ is the average peak position of $n^{th}$ peak of all the 200 recorded spectra. In Fig. \ref{global}, we find that $\Delta \omega$ only shows a fluctuation about its mean value during the measurements. Hence, it is reasonable to conclude that the present data are free from any physical and instrumental artifacts.

Through the above three tests, we nullify any systematic instrumental error or the effect of a slow rise in sample temperature. Furthermore, in our correlation analysis, we include the effect of thermal fluctuations to estimate the extent of correlation. Assuming the effect of temperature fluctuations to be Gaussian,  we have simulated correlation matrices with off-diagonal elements having an independently and identically distributed Gaussian distribution with a standard deviation of 0.05, with a pre-assumed value of the correlation ($\rho$) ranging from $[0,1)$. The $\rho$ is increased by $0.05$, and different matrix norms were calculated to find the extent of correlation.\\

\clearpage
\noindent \textbf{SM9: Raman spectra of KCGO, LNPO, BL-LSMO, and an Organic Compound C$_{12}$H$_{17}$ClN$_4$OS.HCl.}\\

\begin{figure}[h]
\includegraphics[width=0.6\linewidth]{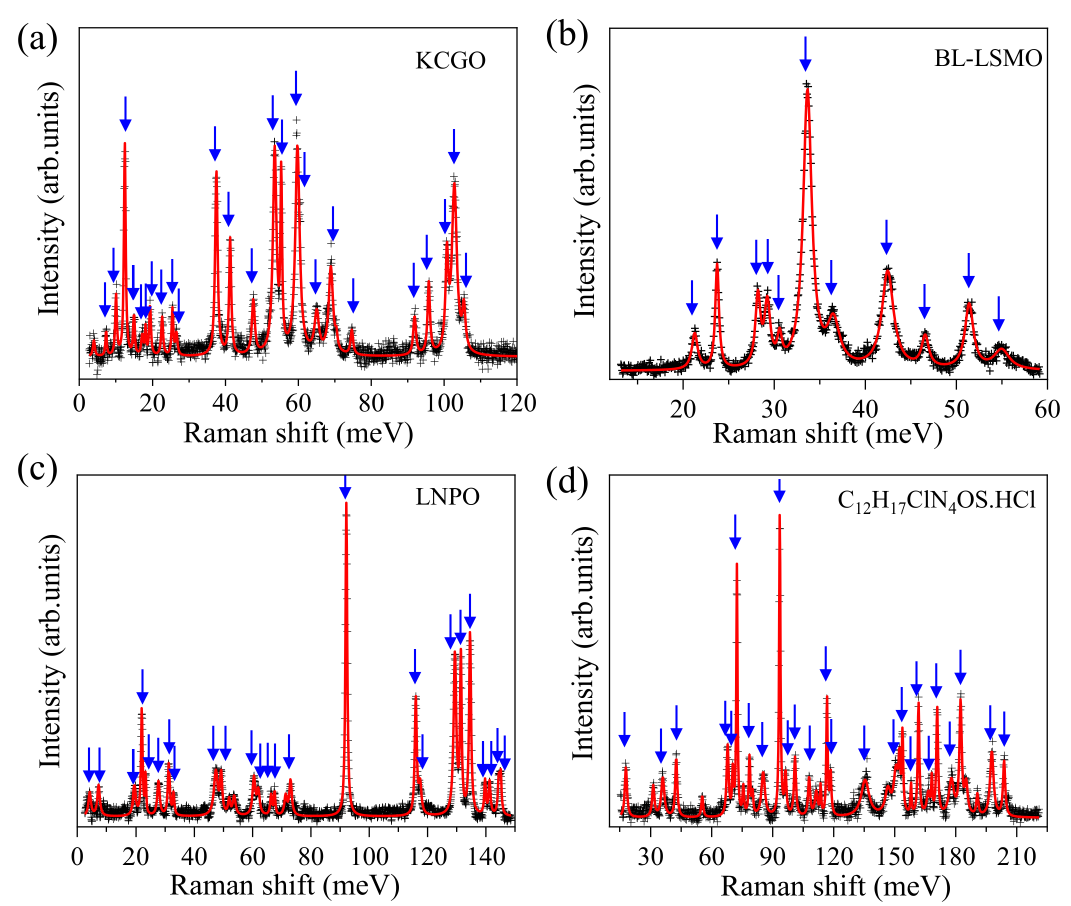} 
\label{other_spectra}
\caption{Raman spectra recorded for (a) KCGO at 80 K,  (b) BL-LSMO at 80 K (c) LNPO at 80 K, and (d) an organic compound. The experimental data are shown in black symbols (+) and the net fitted curve is shown in a red curve. The blue arrows mark the peaks that are considered for the correlation analysis. 25  phonon modes of KCGO, LNPO, and organic compound are considered for correlation analysis. For BL-LSMO, all 11 phonon modes are considered.}
\end{figure}

\noindent \textbf{SM10: Correlation Analysis}\\

In our scientific investigation, the observable  random vector denoted as $\mathbf{X}=(X_1,X_2,\ldots,X_m)^T$, is assumed  to follow a multivariate Gaussian distribution in $\mathbb{R}^m$ for  $m=25 $ (or $m=11$); here $m$ is the Raman modes considered for correlation.  Additionally, we document $n=200$ instances ($n$ is the number of Raman spectra considered for the analysis)  of these observations replicated within the aforementioned experimental conditions. The challenge is this analysis is two folded: (a) Summarizing the magnitude of correlation estimated for $\binom{m}{2}$ pairs (of Raman shift) to a real number, which will be helpful for (b) measuring the significance of deviation from the uncorrelated (null) situation. \\   

\paragraph{Correlation:} From the data, we obtain the observed estimated correlation matrix as 
\begin{eqnarray}
 {\mathbf{R}_{m\times m}}&=& \left(\left(r_{ij}\right)\right)_{i,j} \nonumber\\
&=& \left(\left(Corr(x_i,x_j)\right)\right)_{i,j} \hspace{0.25cm} \forall i,j=1,2,\ldots,m. \label{corr}
\end{eqnarray}
Hence. $ {\mathbf{R}}$ is a random matrix. If the true correlation matrix $\boldsymbol{\varrho}= \left(\left(Corr(X_i,X_j)\right)\right)_{i,j}=\mathbf{I}_m$, then its sample analog $ {\mathbf{R}}$  enjoys a property  that 
\begin{equation}
   r_{ij}=r_{ji}\sim \frac{t}{\sqrt{n-2+t^2}} ~~ \forall i\neq j
   \label{corr_t}
\end{equation}
for independent copies of Students'-$t$ distribution with $(n-2)$ degrees of freedom [72]. \\

\paragraph{Hypothesis testing:}  In the ideal (null) situation, the true correlation matrix $\boldsymbol{\varrho}= \mathbf{I}_m$. But the observed or simulated correlation matrix is a random matrix. A function of the dimension adjusted matrix norm of that random matrix can be used as a statistic to assess its deviation from the identity matrix $\mathbf{I}_m.$ 
Hence, the null distribution of such a statistic can be used to perform a testing of the null hypothesis $H_0:\boldsymbol{\varrho}= \mathbf{I}_m$  against the alternative $H_1:\boldsymbol{\varrho}\neq\mathbf{I}_m.$ Four such statistics have been considered, each of which can map from the space of positive definite symmetric correlation matrices to the real line. We compare the performance of the four test statistics  by their power functions and identify the best possible statistic that will be helpful for establishing the alternative hypothesis more confidently for a common level of significance $\alpha=0.05.$ \\

\paragraph{Test Statistic:}   For  a $m \times m$ sample correlation matrix ${\mathbf{R}}$, we investigate four test statistics obtained from  dimension normalized matrix norms, which include:
\begin{enumerate}
    \item \textbf{Absolute norm} : This matrix norm quantifies the magnitude of the matrix by summing all absolute values of the elements in the matrix, providing a test statistic \\ $$\mathcal{N}_1({\mathbf{R}})=(m(m-1))^{-1}\left(\displaystyle\sum_{i,j=1}^m |r_{ij}|-m\right)$$
     \item \textbf{Frobenius norm }: This matrix norm quantifies the size of the matrix by taking the square root of the sum of the squares of all the elements of the correlation matrix. It leads to a test statistic  \\
     $$\mathcal{N}_2({\mathbf{R}})=(m(m-1))^{-1}\left(\displaystyle\sum_{i,j=1}^m r^2_{ij}-m\right)$$ 
     \item \textbf{One norm}: The one norm of a matrix is defined as the maximum sum of the absolute values of the elements in each column of the matrix, providing a test statistic\\
     $$\mathcal{N}_3({\mathbf{R}})=\displaystyle\max_j (m-1)^{-1}\left(\displaystyle\sum_{i=1}^m |r_{ij}|-1\right)$$
    
     \item \textbf{Spectral range}: The spectral range of the correlation matrix is defined as the difference between the maximum ($\lambda_{\max}$)  and minimum ($\lambda_{\min}$) eigenvalue, and averaged over the dimension \\
     $$\mathcal{N}_4({\mathbf{R}})=(\lambda_{\max}(\mathbf{R})-\lambda_{\min}(\mathbf{R}))/m $$
    
\end{enumerate}

We now conduct a Monte Carlo simulation to explore the distributional characteristics of specific norms related to the estimated correlation matrix  satisfying the condition outlined in Eq. (\ref{corr_t}). We approximated the probability densities of the above four statistics with their kernel densities (smoothed using Gaussian kernel) obtained from the  estimated correlation matrix  ${\mathbf{R}}$ for its $10^6$ replicates maintaining Eq. (\ref{corr}) and Eq. (\ref{corr_t}). From those kernel densities when $m=25$ it is observed that the distributions are positively skewed and their $95th$ quantiles  are computed respectively as \\

\begin{center}
$\mathcal{Q}_{1,95}=0.02318865$, $\mathcal{Q}_{2,95}=0.02900500$, 
$\mathcal{Q}_{3,95}=0.03646224$, 
$\mathcal{Q}_{4,95}=0.02264022$.
\end{center}

\begin{figure} [ht]
\includegraphics[width=0.45\linewidth]{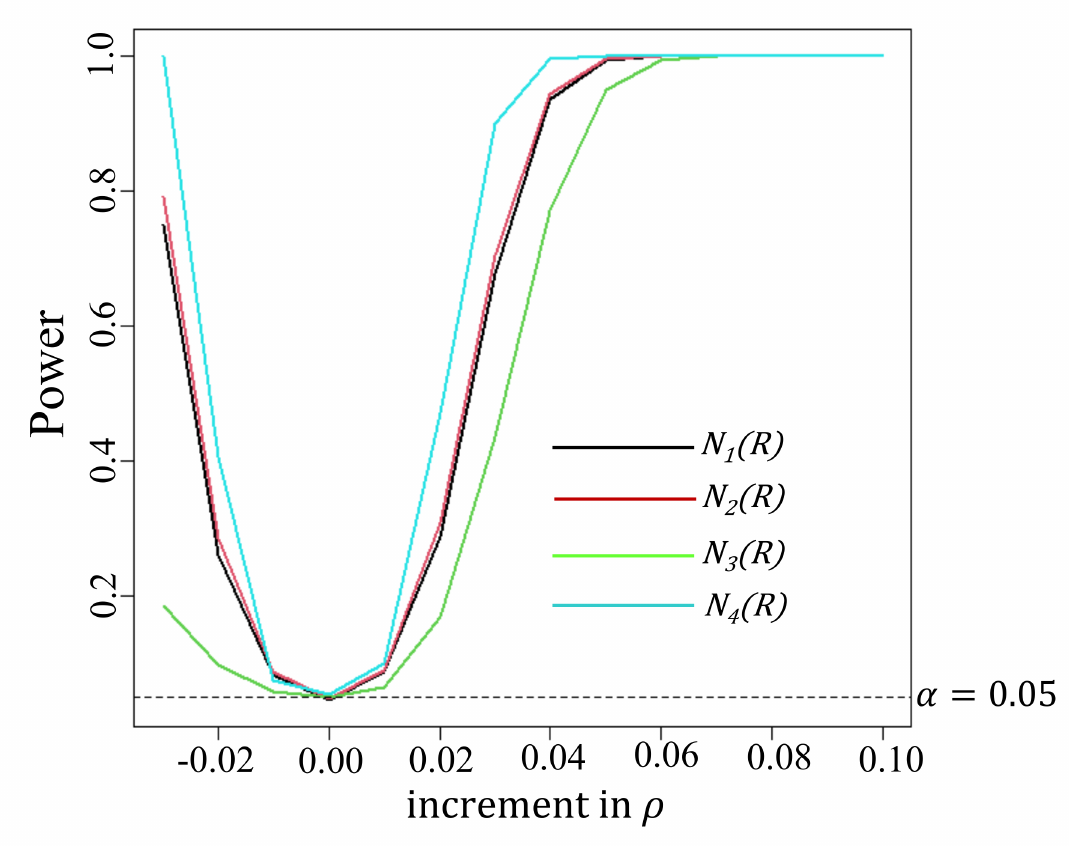}
\caption{Power Comparisons of different matrix norms.}
\label{power_comparision}
\end{figure}

\paragraph{Magnitude assessment:}As an alternative situation ($H_1$), we have simulated a form of a multivariate Gaussian distribution with a zero mean and an equi-covariance matrix 
\begin{equation}
    \varSigma_{m \times m}= ((1 \mathbf{1}_{\{i=j\}}+ \rho \mathbf{1}_{\{i\neq j\}}))_{i,j}
\end{equation}
For different values of $\rho\in[0,1)$, the dimension adjusted matrix norms are then computed for the simulated data (\ref{power_comparision}).   The matrix norms of these simulated matrices are then approximated as a polynomial function of correlation ($\rho$). Finally, a reverse mapping of the approximated polynomial is performed to obtain a one-point correlation as the inverse image of the norm of the correlation matrices derived from the experimental data.
In supplementary Fig \ref{power_comparision}, it is interesting to observe that a small deviation from the null situation  ($\boldsymbol{\varrho}=\mathbf{I}_m$ and Eq. (\ref{corr_t})) is sharply captured by all four statistics defined above through an upper tail test in each case. Moreover, the statistic $\mathcal{N}_4$  provides the maximum and dominant power among those, as it is most sensitive to changes in the value of $\rho$. In the context of the present work, it indicates that the spectral range $\mathcal{N}_4$ of the phonon–phonon correlation matrix is expected to yield the most sensitive measure of collective lattice dynamics. 
In this statistical analysis, the power of a test represents the probability of rejecting the null hypothesis in favor of the alternative at pre-specified level $\alpha=0.05$  when the alternative hypothesis is true. This indicates the sensitivity of the test in measuring the deviation from the null hypothesis; whereas the level (dashed line) is the same probability when the null hypothesis is true.  
\\

Now we simulated the correlation matrix multivariate Gaussian distribution with zero mean ($\rho$ =0) and estimated the above dimension normalized norms. The same has been repeated with an increment of 0.05 to cover the interval $[0,1)$.  Also, note that maintaining $\varSigma_{m \times m}$ as a positive definite matrix, the effective range of $\rho$ is [-0.04, 1) for $m =25$ and [-0.09, 1) for $m = 11$. At each instance, all four test statistics have been calculated. Supplementary Fig. \ref{correlation} shows the quadratic polynomial fit to the dimension normalized matrix norms obtained from varying  $\rho$ values for $m=25$ and $11$. The red region in \ref{correlation} represents extrapolated matrix norm values for higher $\rho$ values.
Finally, a reverse mapping of the approximated polynomial is performed to obtain a one point correlation as an inverse image of the norm of the correlation matrices obtained from experimental data. 
For example, the $\mathcal{N}_1({\mathbf{R}})$ value of NCGO at 80 K is 0.77191.  The corresponding extent of correlation $\rho$=0.779 has been obtained. The same is illustrated in \ref{correlation} (a). By following a similar procedure, we first calculated the different norm values experimentally, and then, by reverse mapping, we obtained the extent of correlation for all five compounds (six sets). 
Supplementary Table \ref{co_table} presents the experimental norm values; these are compared with the simulated norm values (from \ref{correlation}) to assess the extent of correlation.

\newpage
\begin{figure}[h]
\includegraphics[width=0.85\linewidth]{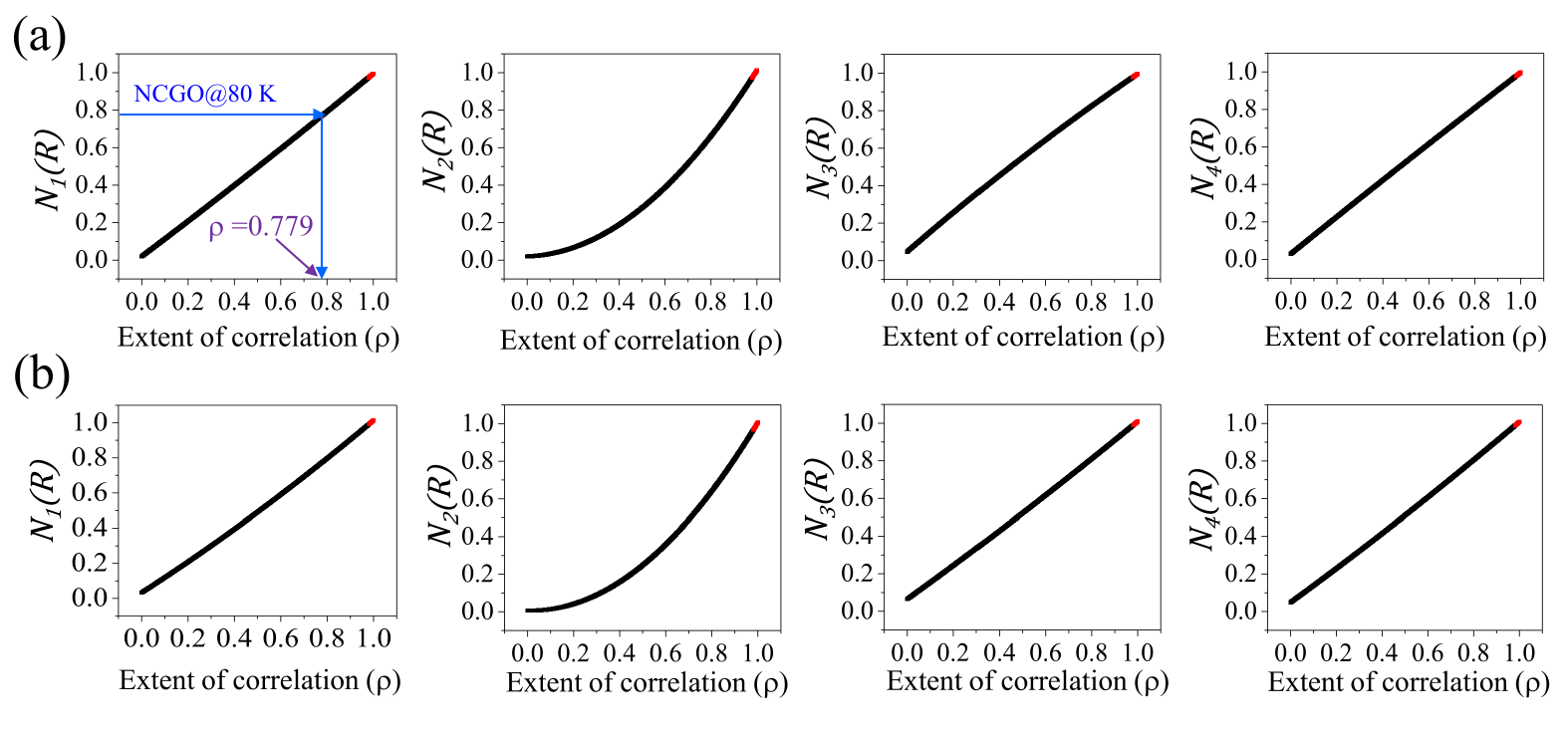} 
\caption{Variation of different norm values with the extent of correlation for (a) $m = 25$ and (b) $m = 11$. Here, the red-marked region is the extrapolated region for the polynomial fit. As an example, the first panel in (a) provides the procedure followed to estimate the extent of correlation ($\rho$), described in the main text for the spectra recorded at 80 K.  From the experimental data $\mathcal{N}_1({\mathbf{R}})$ value of NCGO at 80 K is estimated as 0.77191 (horizontal arrow).  The corresponding extent of correlation $\rho$ = 0.779 has been obtained (vertical arrow). The ranges of values of $\rho$ in Table  1 of the main text are obtained spread of values using all four norms.}
\label{correlation}
\end{figure}

\begin{table}[h]
\caption{Estimated different matrix norms for five compounds (six data sets)}
\centering
\scriptsize

\begin{tabular}{cccc}

\hline
\hline
Compound & Norms & Experimental  & Extent of correlation \\
& & values & \\
\hline
& $\mathcal{N}_1$ &0.77191&0.779\\
& $\mathcal{N}_2$ &0.66425&0.802\\
NCGO at 80 K& $\mathcal{N}_3$ &0.88239&0.867\\
& $\mathcal{N}_4$ &0.81888&0.812\\
\hline

& $\mathcal{N}_1$ &0.17773&0.167\\
& $\mathcal{N}_2$ &0.05882&0.181\\
NCGO at 400 K& $\mathcal{N}_3$ &0.30678&0.267\\
&$\mathcal{N}_4$ &0.22296&0.202\\
\hline

& $\mathcal{N}_1$ &0.19662&0.188\\
& $\mathcal{N}_2$ &0.07297&0.215\\
KCGO at 80 K& $\mathcal{N}_3$ &0.35365&0.298\\
& $\mathcal{N}_4$ &0.25463&0.228\\
\hline
\end{tabular}
\hspace{2em}
\begin{tabular}{cccc}
\hline
\hline
Compound & Norms & Experimental  & Extent of correlation\\
& & values & \\
\hline

& $\mathcal{N}_1$ &0.20870&0.201\\
& $\mathcal{N}_2$ &0.07028&0.264\\
BL-LSMO at 80 K& $\mathcal{N}_3$ &0.30527&0.268\\
&$\mathcal{N}_4$ &0.25978&0.233\\
\hline

& $\mathcal{N}_1$ &0.17010&0.161\\
&$\mathcal{N}_2$ &0.050421&0.158\\
LNPO at 80 K& $\mathcal{N}_3$ &0.30066&0.249\\
& $\mathcal{N}_4$ &0.21298&0.185\\
\hline

& $\mathcal{N}_1$ &0.16511&0.155\\
& $\mathcal{N}_2$ &0.04836&0.152\\
Organic compound & $\mathcal{N}_3$ &0.27067&0.215\\
& $\mathcal{N}_4$ &0.22436&0.197\\

\hline
\label{co_table}

\end{tabular}
\end{table}

\end{document}